\documentclass[aps,prb,reprint,nofootinbib]{revtex4-2}

\usepackage{amsmath,amssymb,bm}
\usepackage{graphicx}
\usepackage{booktabs}
\usepackage{xcolor}
\usepackage{hyperref}

\newcommand{\MnSi}{Mn$_5$Si$_3$}
\newcommand{\q}{\mathbf q}
\newcommand{\T}{\mathbf T}
\newcommand{\G}{\Gamma}
\newcommand{\nhat}{\hat{\mathbf n}}
\newcommand{\e}{\mathbf e}
\newcommand{\Lop}{\mathbf L}
\newcommand{\m}{\mathbf m}
\newcommand{\M}{\mathbf M}

\begin{document}

\title{Landau theory and exchange instabilities in Mn$_5$Si$_3$: A case against strain-induced altermagnetism}

\author{K. D. Belashchenko}
\affiliation{Department of Physics and Astronomy and Nebraska Center for Materials and Nanoscience, University of Nebraska--Lincoln, Lincoln, Nebraska 68588, USA}

\date{\today}

\begin{abstract}
Thin-film Mn$_5$Si$_3$ is one of the most studied altermagnetic candidates thanks to its metallicity, demonstrated anisotropic anomalous transport properties, and proposed $d$-wave exchange splitting pattern that would enable spin-polarized transport and various spintronic applications. Its postulated altermagnetic structure has zero propagation vector, in contrast to the collinear antiferromagnetic bulk phase (AFM2) which orders at the $M$ star.
In this work, the two phases are analyzed using Landau theories, first-principles calculations of the paramagnetic instabilities, and Monte Carlo simulations. AFM2 appears in a Landau theory as a symmetry-protected inversion-even, permutation-odd mode at a single arm of the $M$ star. At $\Gamma$, the same intracell ordering pattern belongs to the collinear branch of an $E_{2g}$ order parameter. In both cases, higher-order terms are required for the phase selection. First-principles calculations for the paramagnetic, disordered-local-moment state correctly identify the leading exchange instability at the $M$ star, and the resulting classical Heisenberg model orders at the bulk experimental temperature scale into the orthogonal $3M$ phase favored by single-site entropy.
Magnetostructural coupling to $E_{2g}$ deformations favors the single-$M$ AFM2 phase. The estimated homogeneous-strain contribution to the relevant quartic coefficient is only about a quarter of the competing single-site entropic term near the bulk transition temperature, but internal $E_{2g}$ relaxations are expected to provide additional stabilization.
The $\Gamma$-point $E_{2g}$ mode, whose Landau theory contains the altermagnetic sector, is substantially weaker and further suppressed by epitaxial strain comparable to that reported for \MnSi/Si(111) films exhibiting anomalous transport. The same strain reduces the leading magnetic exchange scale. These results provide a natural explanation for the bulk $M$-point instability but strongly disfavor the relocation of the propagation vector from $M$ to $\G$ in a moderately strained bulklike \MnSi\ film, leaving the identity of the Hall-active phase unresolved.
\end{abstract}

\maketitle

\section{Introduction}

Hexagonal \MnSi\ exhibits a highly unusual magnetic order involving both crystallographic symmetry lowering and partial magnetic order \cite{BrownForsythNunezTasset1992,BrownForsyth1995}. It has  recently attracted intense attention due to the unconventional transport properties observed in thin films
\cite{Reichlova2024,Kounta2023,Leiviska2024,Han2024SciAdv,Badura2025,Han2025PRA,Rial2024variants,han2025,Skobjin2026,mencos2025}. These observations have been attributed to a collinear altermagnetic phase which does not exist in the bulk but is obtained from the collinear antiferromagnetic bulk phase (denoted AF2 or AFM2) by relocating the ordering vector from the zone-boundary $M$ point to the zone center \cite{Reichlova2024}. This altermagnetic phase would make \MnSi\ a rare example of a metallic $d$-wave altermagnet with compelling applications in spintronics \cite{Smejkal1,Smejkal2,Song_review_2025,Jungwirth2026}.

The AFM2 phase itself is rather exotic \cite{BrownForsyth1995}. The paramagnetic space group $P6_3/mcm$ includes Mn atoms in a $4d$ orbit (Mn1) and a $6g$ orbit (Mn2). Even without spin-orbit coupling, the phase transition into AFM2 simultaneously breaks time-reversal and space-group symmetry down to orthorhombic $Ccmm$, and, most peculiarly, involves only two thirds of the Mn2 sites, while the remaining third, along with all the Mn1 sites, remains completely disordered. In first-principles calculations, the AFM2 phase was modeled assuming the disordered Mn2 sites have a vanishing local moment and found to be locally stable \cite{dosSantos2021}. However, the thermodynamic origin of the AFM2 phase and the microscopic interactions promoting its selection from the paramagnetic state have not been established.

In contrast to the well-established bulk AFM2 phase, the proposed altermagnetic phase in thin \MnSi\ films has been inferred indirectly from anomalous transport, its variant structure, and lateral spin injection \cite{Reichlova2024,Rial2024variants,mencos2025}; there is neither a direct observation of the ordering vector nor a microscopic explanation for its proposed switch from $M$ to $\G$-point. The anomalous Hall (AHE) and Nernst (ANE) effects observed in thin films persist above 200 K, more than twice the bulk AFM2 ordering temperature \cite{Reichlova2024,Badura2025,Han2024SciAdv,Han2025PRA}; such an enhancement of a N\'eel temperature in a modestly strained film retaining the bulk structure is highly unusual.
Further, finite AHE and ANE in the proposed altermagnetic phase require additional symmetry lowering, such as a low-symmetry orientation of the N\'eel vector \cite{Reichlova2024} or a spontaneous monoclinic distortion of the Mn2 site environment \cite{Rial2024variants}, neither of which has been directly confirmed.

Thus, the magnetic state responsible for the anomalous transport in thin \MnSi\ films remains uncertain, as does the mechanism that stabilizes it. One possibility is that epitaxial strain favors a magnetic ordering competing with bulk AFM2, whether it is the proposed $\Gamma$-point altermagnetic phase or some other ordering. Alternatively, this state may be associated with a modified composition or structure. It is worth noting that \MnSi-type compounds accommodate interstitial dopants on the empty $2b$ sites at the centers of Mn2 octahedra \cite{Corbett1998}; carbon intercalation in \MnSi, for example, drastically changes its magnetic properties \cite{Gajdzik2000}. The sixfold enhancement of the anomalous Nernst conductivity observed in \MnSi/Al$_2$O$_3$ (0001) films with 2\% excess Mn \cite{Han2025PRA} further illustrates the sensitivity of anomalous transport to composition.

Using a combination of Landau theory, first-principles calculations within the disordered-local-moment (DLM) approach, and Monte Carlo simulations, this paper clarifies the origin of the bulk AFM2 phase and examines whether moderate homogeneous strain can favor a competing magnetic instability or enhance the magnetic ordering scale in bulklike \MnSi. This analysis shows that an instability at $\Gamma$ is strongly disfavored compared to the $M$ point (and other in-plane zone-boundary modes) and is further suppressed by tensile epitaxial strain typical for thin \MnSi\ films grown on Si(111). Moreover, the same strain reduces the leading magnetic exchange scale rather than enhancing it. These results suggest that the anomalous-Hall state in such films requires additional microscopic ingredients.

The paper is organized as follows. Section \ref{sec:landau} constructs Landau theories for the bulk $M$-point AFM2 ordering (Subsection \ref{subsec:Mlandau}) and for the altermagnetic order (Subsection \ref{subsec:Glandau}). Section \ref{sec:dlm} examines the quadratic ordering tendencies in the paramagnetic state using the DLM approach.
Section \ref{sec:MC} describes the results of Monte Carlo simulations. Section \ref{sec:magnetoelastic} estimates the magnetoelastic coupling of the $M$-star ordering to symmetry-matched in-plane shear strain, which may be relevant to the phase selection. The results are discussed in Section \ref{sec:discussion}, and the conclusions are drawn in Section \ref{sec:conclusions}.

\section{Landau theories of the magnetic ordering in {M\MakeLowercase{n}$_5$S\MakeLowercase{i}$_3$}}
\label{sec:landau}

\subsection{General symmetry considerations}
\label{subsec:GeneralSymmetry}

On cooling below $T_{N2}\approx100$ K from the paramagnetic $P6_3/mcm$ phase, bulk \MnSi\ first enters the AFM2 phase with a doubled magnetic unit cell and an $M$-point propagation vector, in which two thirds of the Mn2 sites order antiparallel to each other, while the remaining Mn2 and all the Mn1 sites remain disordered \cite{BrownForsyth1995}. The $T_{N2}$ transition appears to be first-order but only weakly; we therefore assume that its key features must follow from an appropriate Landau theory.
At a lower temperature $T_{N1}\approx66$ K the system enters a noncollinear, noncoplanar AFM1 phase \cite{BrownForsythNunezTasset1992} involving both Mn2 and Mn1 atoms, which exhibits a large topological Hall effect \cite{Surgers2014,Surgers2024}. The magnetic structure of this phase is not fully established \cite{Biniskos2022,Surgers2024} and will not be the focus here. 

\begin{figure}[htb]
    \centering
    \includegraphics[width=0.85\linewidth]{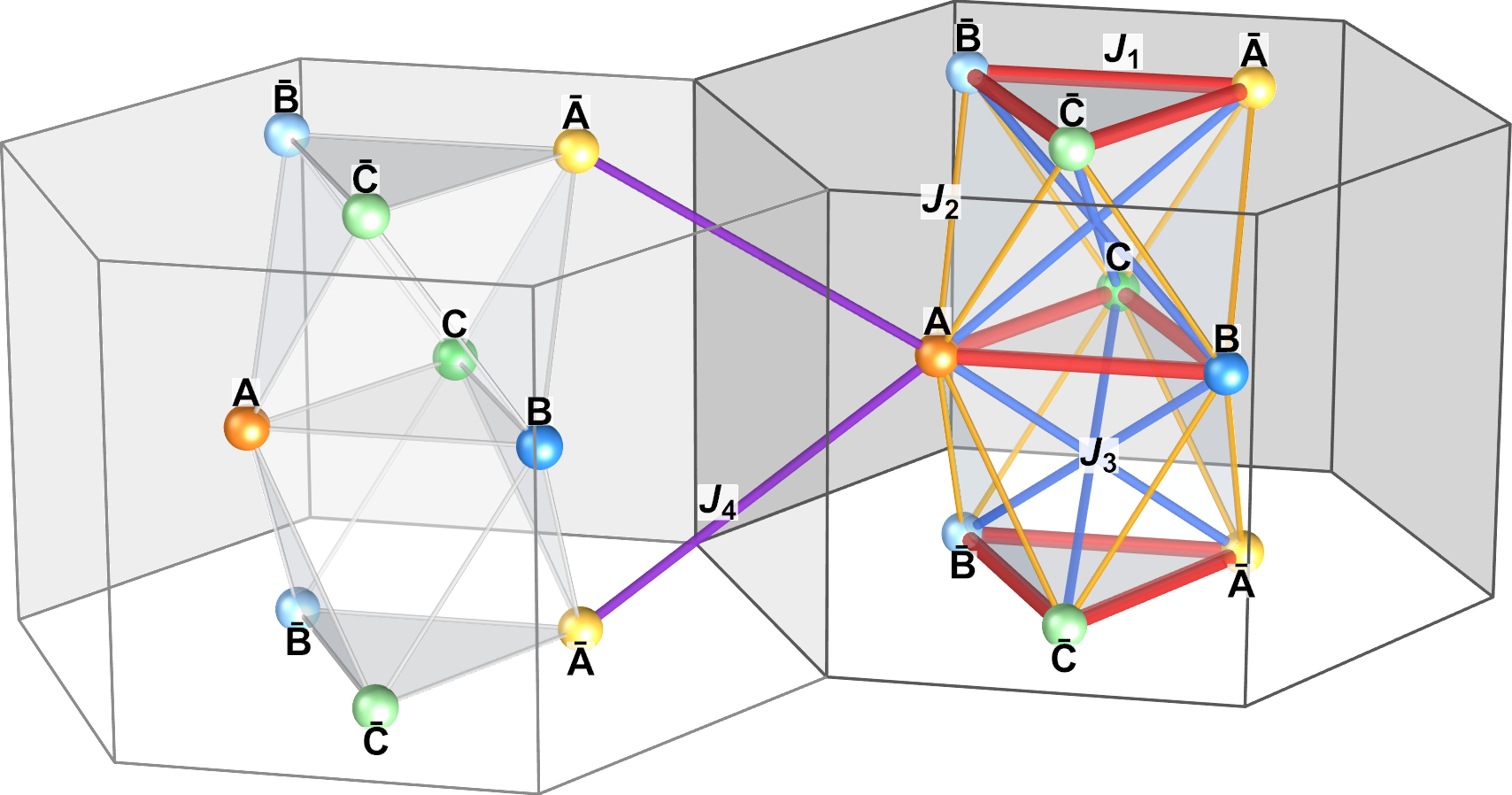}
    \caption{Two neighboring polyhedral primitive cells (Wigner-Seitz cells translated by $\mathbf{c}/4$) of \MnSi\ showing only the $6g$ orbits of Mn2 atoms, which are labeled $A$, $\bar A$, $B$, $\bar B$, $C$, $\bar C$, where $X$ and $\bar X$ are inversion-related pairs. Four nearest exchange bonds are labeled as $J_1,\dots,J_4$. Equivalent sets of bonds are highlighted in the same colors.}
    \label{fig:Jsketch}
\end{figure}

The Mn2 orbit, which is illustrated in Fig. \ref{fig:Jsketch}, forms nearly octahedral clusters, with opposite triangular faces parallel to the $(001)$ plane and a slight trigonal elongation along the $c$ axis. These clusters stack along $c$ into face-sharing chains, with two clusters per primitive cell. Viewed as structural units, the octahedral chains are arranged in a triangular lattice; its dual honeycomb lattice is occupied by straight chains of Mn1 atoms (not shown). Each Mn1 atom has six equivalent Mn2 neighbors, one from each sublattice of the $6g$ orbit. The center of each Mn2 octahedron is an inversion center, and the six Mn2 sublattices are labeled as inversion-related pairs ($A$, $\bar A$), ($B$, $\bar B$), and ($C$, $\bar C$), where $ABC$ or $\bar A\bar B\bar C$ sites in the same Wigner-Seitz cell form equilateral triangles in a (001) plane. The polyhedral cells shown in Fig. \ref{fig:Jsketch} are obtained by translating inversion-centered Wigner–Seitz cells by $\mathbf c/4$. 

The lattice translations are defined as $\mathbf a_1=(\sqrt{3}/2,-1/2,0)a$, $\mathbf a_2=(\sqrt{3}/2,1/2,0)a$, and $\mathbf a_3=c(0,0,1)$, 
and the $6g$ basis is assigned, within the inversion-centered Wigner-Seitz cell at the origin, as
$\mathbf{r}_A = (-x,0,1/4)$, $\mathbf{r}_B = (x,-x,1/4)$, $\mathbf{r}_C = (0,x,1/4)$, and $\mathbf{r}_{\bar X}=-\mathbf{r}_{X}$. For later reference, the reciprocal lattice vectors are $\mathbf G_1=(2\pi/a)(1/\sqrt{3},-1,0)$, $\mathbf G_2=(2\pi/a)(1/\sqrt{3},1,0)$, and $\mathbf G_3=(2\pi/c)(0,0,1)$.

For a given ordering vector $\mathbf q$, the quadratic part of the Landau free energy can be written as
\begin{equation}
F_2=\frac{1}{2}\sum_{\alpha\beta}
\m_\alpha(-\mathbf q)\cdot
\mathcal K_{\alpha\beta}(\mathbf q)
\m_\beta(\mathbf q),
\label{F2}
\end{equation}
where $\alpha,\beta\in\{A,\bar A, B, \bar B, C, \bar C\}$, $\mathcal K(\mathbf q)$ is the (Hermitian) inverse susceptibility matrix, and the Fourier transformed ordering amplitudes are defined as  $\m_\alpha(\q)=\sum_\T \m_\alpha(\T)\exp(i\q\T)$ with lattice translations $\T$. In a microscopic Heisenberg description, $\mathcal K(\mathbf q)$ is related to the Fourier-transformed exchange matrix. The leading magnetic instability corresponds to the eigenmode of $\mathcal K(\mathbf q)$ whose eigenvalue first vanishes on cooling.

The little space group $G_{\q}$ is the subgroup of the full space group $G$ that leaves $\q$ invariant up to a Bragg vector. 
The symmetry constraints on $\mathcal K_{\alpha\beta}(\q)$ are set by the projective representation $D(\q)$ of the corresponding little cogroup $\bar G_\q$. The matrices $D_g(\q)$ act on the ordering amplitudes $\m_\alpha(\q)$. In the nonrelativistic limit considered here, thanks to the global spin-rotation symmetry, we only need to consider the action of the group elements $g\in G_{\q}$ on the lattice and not on the spin degrees of freedom. This action takes a basis site at $\mathbf r_\alpha$ to a site $\mathbf r_{P_g\alpha} + \T_\alpha(g)$, where $P_g\in S_6$ is a permutation of the six sublattices, and $\T_\alpha(g)$ is a lattice translation. The set of $P_g$ and $\T_\alpha(g)$ defines the representation matrices
\begin{equation}
(D_g)_{\beta\alpha}=(P_g)_{\beta\alpha}\exp\left[i\q \T_\alpha(g)\right].
\label{symmetryK}
\end{equation}
The matrices $D_g$ for pure translations contain simple Fourier phases that cancel in Eq. \eqref{F2} and impose no restrictions on $\mathcal K(\q)$, and the nonsymmorphic components of $g\in G_\q$ are already accounted for by $\T_\alpha(g)$. Therefore, to determine the symmetry of $\mathcal K(\q)$, we should generally treat Eq. \eqref{symmetryK} as a projective representation of the little cogroup $\bar G_\q$ corresponding to $G_{\q}$, i.e., it must obey
\begin{equation}
    D_{g}^\dagger \mathcal K(\q)D_{g}=\mathcal K(\q)
    \label{eq:Kinvariance}
\end{equation}
where $g\in G_\q$ is a representative space-group element corresponding to $\bar g\in \bar G_\q$.

At $\Gamma$, the point group represented by $D_g(\Gamma)$ is a six-color  point group \cite{Harker1981,Lifshitz-RMP}, which is a direct generalization of an antisymmetry point group to a six-sublattice crystal. At a finite wave vector $\mathbf q$, the induced sublattice permutations are decorated by phase factors associated with the lattice translations needed to return the transformed sites to the chosen primitive cell.

We are interested in Landau theories that can generate the AFM2 ordering pattern, in which the ordering amplitudes are opposite on two effective sublattices and the third one vanishes. We will see that such orderings can naturally occur at the $\Gamma$ and $M_\nu$ points, where $\nu$ denotes one of the three arms of the $M$ star. 
Because these points are invariant under inversion, the corresponding eigenmodes of $\mathcal K(\Gamma)$ and $\mathcal K(M_\nu)$ may be classified by inversion parity $p=\pm1$, such that $\m_{\bar X}=p\m_{X}$ for $X\in\{A,B,C\}$. We restrict our consideration to the parity-even sector, which is relevant for both the bulk $M$-point AFM2 phase and for the hypothetical altermagnetic $\Gamma$-point phase. Thus, we are left with three even ordering amplitudes $\m^+_A$, $\m^+_B$, $\m^+_C$ and the corresponding three-dimensional inversion-even representation $D^+(\Gamma)$ or $D^+(M_\nu)$.

As explained below, the $\mathcal K(\Gamma)$ and $\mathcal K(M_\nu)$ eigenmodes describing AFM2-like phases exert no bilinear exchange field on the Mn1 atoms. This implies that the Landau free energy can have no bilinear couplings between these Mn2 eigenmodes and the Mn1 ordering amplitudes. We therefore exclude Mn1 atoms from the Landau theories below, while noting that interaction mediated by Mn1 may, in principle, renormalize higher-order terms in the Landau free energy.

\subsection{$M$-star ordering}
\label{subsec:Mlandau}

We start with the case when the leading magnetic instability of the paramagnetic phase occurs at the $M$ star, which is  directly relevant to the experimental AFM2 structure obtained from neutron diffraction \cite{BrownForsyth1995}.

Let $M_a$, $M_b$, and $M_c$ be the three arms of the $M$ star. A key observation is that once an arm is chosen, the equivalence of the three effective $A$, $B$, $C$ sublattices is lost. Indeed, the (orthorhombic, $mmm$) little cogroup of a given $M_\nu$ arm does not contain threefold rotations, which cyclically permute the $A$, $B$, and $C$ sublattices, but retains symmetry planes that interchange two of them.
It is convenient to fix the notation so that sublattices $B$ and $C$ are equivalent at $M_a$, $C$ and $A$ at $M_b$, and likewise $A$ and $B$ at $M_c$. In the crystallographic setting specified in Section \ref{subsec:GeneralSymmetry}, this assigns $M_a=\mathbf{G}_2/2$, $M_b=(\mathbf{G}_1+\mathbf{G}_2)/2$, and $M_c=\mathbf{G}_1/2$.

Because $M_b$ lies on the $k_x$ axis, the little cogroup $\bar G_{M_b}$ is $m_xm_ym_z$. All phase factors reduce to unity, and in the space of the three even ordering amplitudes $\m^+_X$, we find $D^+_{m_z}(M_b)=\hat1$, and $D^+_{m_x}(M_b)=D^+_{m_y}(M_b)=S_{AC}$. Thus, sublattice $B$ is distinguished at $M_b$. According to Eq. \eqref{eq:Kinvariance}, the only restriction on $\mathcal K(M_b)$ is that it commutes with $S_{AC}$, which leads to a generic form
\begin{equation}
     \mathcal K(M_b)=\begin{pmatrix}
\kappa_1 & \kappa_3 & \kappa_2 \\
\kappa_3 & \kappa_4 & \kappa_3 \\
\kappa_2 & \kappa_3 & \kappa_1
\end{pmatrix},
\end{equation}
where all parameters are real because $\mathcal K(\q)=\mathcal K^\ast(-\q)$ and $-\M_b=\M_b+\mathbf G$.
This matrix generally has three nondegenerate eigenvectors, two of which are even under $S_{AC}$ and one odd $(-1,0,1)$; the latter is fixed by permutation symmetry. A nondegenerate eigenvector describes \emph{collinear} ordering with an arbitrary orientation in spin space. The permutation-odd mode at a single arm of the $M$ star is exactly the experimentally observed AFM2 phase. Of course, vanishing magnetization on one sublattice does not mean that its local moment disappears in this instability; its orientation simply remains disordered.

The three arms of the $M$ star are related by threefold rotation, and the AFM2-like eigenmodes are:
\begin{align}
\label{eq:Ma_pattern}
M_a&: \quad (A,B,C) \propto (0,1,-1),\\
\label{eq:Mb_pattern}
M_b&: \quad (A,B,C) \propto (-1,0,1),\\
\label{eq:Mc_pattern}
M_c&: \quad (A,B,C) \propto (1,-1,0).
\end{align}
We introduce (real) spin-space vector order parameters $\Lop_\nu$, $\nu\in\{a,b,c\}$, representing spatial ordering patterns \eqref{eq:Ma_pattern}-\eqref{eq:Mc_pattern}.
Explicitly, for a primitive cell at $\T=n_1\mathbf a_1+n_2\mathbf a_2+n_3\mathbf a_3$, the sign factors
\begin{equation}
\eta_a=(-1)^{n_2}, \quad
\eta_b=(-1)^{n_1+n_2}, \quad
\eta_c=(-1)^{n_1}
\label{eq:etas}
\end{equation}
define the local sublattice ordering amplitudes for a general ordering pattern involving the whole $M$ star:
\begin{align}
\m^+_A(\T)&=\eta_c \Lop_c-\eta_b \Lop_b,\nonumber\\
\m^+_B(\T)&=\eta_a \Lop_a-\eta_c \Lop_c,\label{eq:Mstar_realspace}\\
\m^+_C(\T)&=\eta_b \Lop_b-\eta_a \Lop_a.\nonumber
\end{align}
A single nonzero $\Lop_\nu$ gives the experimental AFM2 phase. Combinations of two or three $\Lop_\nu$ describe multi-$M$ states that may be collinear or noncollinear.

Whether ordering occurs at a single $M$ arm or as a superposition of multiple arms depends on the quartic terms in the Landau free energy, which maps to the generic case of an $M$-star $O(3)$ order parameter considered in Ref. \cite{Jin2025}.
Let $Q_{\mu\nu}=\Lop_\mu\cdot\Lop_\nu$ and let $Q_d$ and $Q_\times$ be the diagonal and off-diagonal parts of $Q$. Further, let $q_0=\frac13Q_{\mu\mu}$ and $Q_\Delta=Q_d-q_0I$, so that $Q=q_0\hat1+Q_\Delta+Q_\times$. To quartic order, the free energy $F_M$ for the $M$-star order parameter is \cite{Jin2025}:
\begin{equation}
F_M = a_M Q_{\mu\mu}
+u_M (Q_{\mu\mu})^2
+v_M\mathop{\mathrm{Tr}} Q_\Delta^2
+w_M\mathop{\mathrm{Tr}} Q_\times^2+\cdots,
\label{eq:FM_landau}
\end{equation}

\begin{figure}
    \centering
    \includegraphics[width=0.85\linewidth]{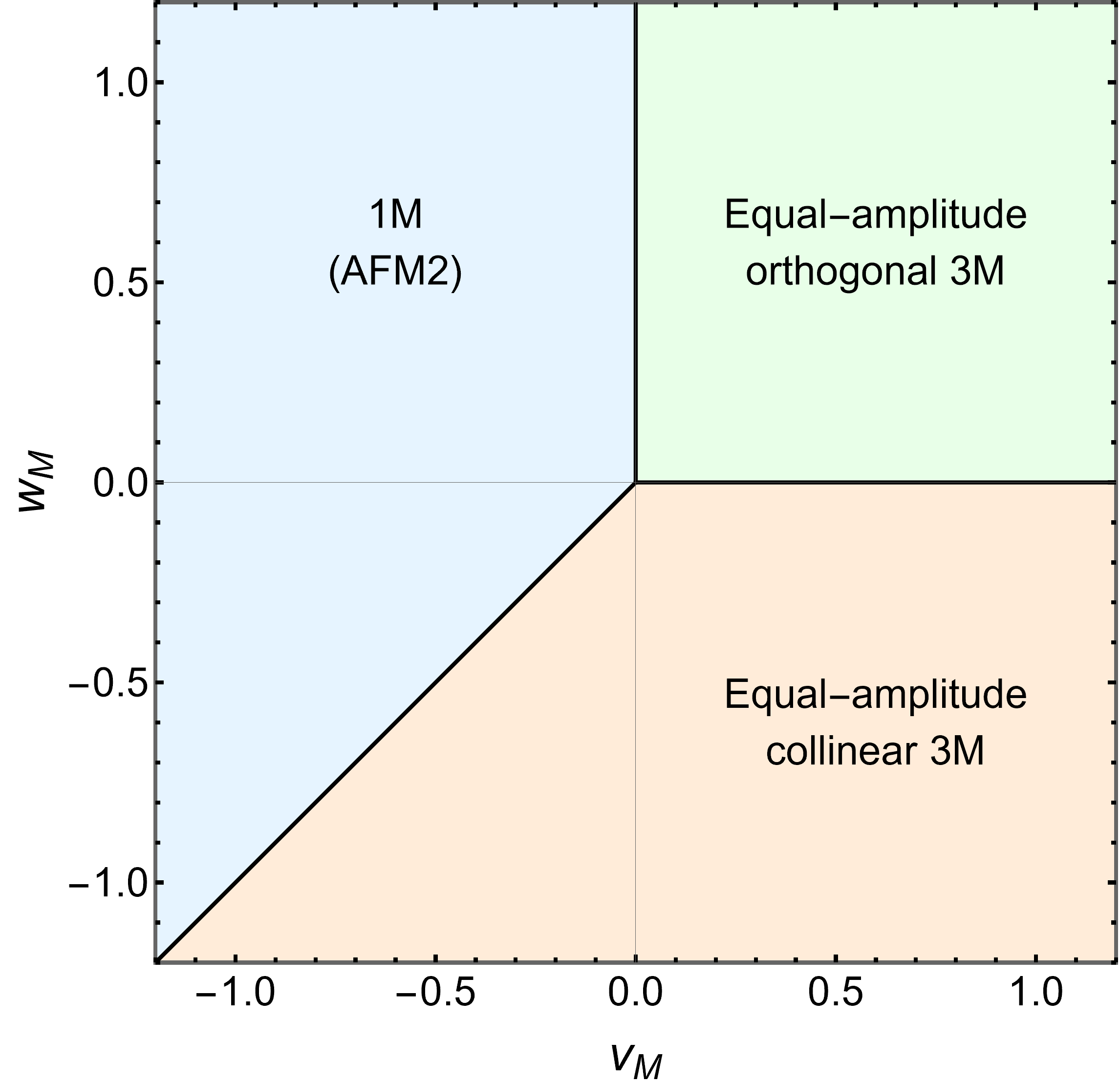}
    \caption{Phase diagram of the $M$-point ordering in the permutation-odd sector. It is assumed that the quartic part of the free energy is positive definite. The phase selection is controlled by the quartic parameters $v_M$ and $w_M$ in the free energy \eqref{eq:FM_landau}, shown here in arbitrary units.}
    \label{fig:MPD}
\end{figure}

The quadratic term with $a_M$ and the quartic term with $u_M$ depend only on the total amplitude $Q_{\mu\mu}=\sum_\mu \Lop_\mu^2$. At a fixed $Q_{\mu\mu}$, 
$v_M>0$ prefers $\Lop_\mu^2$ to be equal for all three $M_\nu$ arms, and $v_M < 0$ favors ordering at a single arm. The $w_M$ term comes with the inter-arm dot products in $\mathop{\mathrm{Tr}}Q^2_\times=2\left[(\Lop_a\cdot\Lop_b)^2+(\Lop_b\cdot\Lop_c)^2+(\Lop_c\cdot\Lop_a)^2\right]$ and is the only one that depends on the relative orientations of $\Lop_\nu$, favoring orthogonal configurations at $w_M>0$ and collinear ones at $w_M<0$.
The phase diagram below $T_c$ is shown in Fig.\ \ref{fig:MPD}. The experimentally observed single-$M$ AFM2 phase occurs for $v_M<0$ and $w_M>v_M$. 

In the classical rigid-spin Heisenberg model, mean-field theory leads to a purely entropic quartic free-energy term
$F_{\rm ent}^{(4)} = bT\sum_{i}\m_i^4$, where $b=9k_B/20>0$, and the summation is over all magnetic sites. The resulting contribution to the quartic free-energy terms depends on how the $M$-star order parameter is embedded in the lattice. Substitution of \eqref{eq:Mstar_realspace} and \eqref{eq:etas} followed by averaging over the $\eta_\nu$ phases gives $u^\mathrm{ent}_M=\frac65k_BT$ per primitive cell, $v^\mathrm{ent}_M=\frac34u^\mathrm{ent}_M>0$, and $w^\mathrm{ent}_M=2v^\mathrm{ent}_M$. This would place the system into the equal-amplitude orthogonal $3M$ phase rather than the experimentally observed single-arm AFM2 phase. Thus, stabilization of the AFM2 phase requires additional interactions or correlations that are not captured by the Heisenberg model in the mean-field approximation.

Apart from the non-Heisenberg electronic interactions, there is a natural magnetoelastic mechanism that can contribute to the quartic terms in \eqref{eq:FM_landau}. Spin-rotation-invariant magnetic bilinears $Q_{\mu\nu}$ can couple linearly to symmetry-compatible lattice deformation modes. The two independent components of $Q_\Delta$ carry zero wave vector, transform as the $E_{2g}$ irreducible representation (irrep) of the $6/mmm$ point group, and can couple to lattice deformation modes transforming under the same irrep, including in-plane shear strain components ($\varepsilon_{xx}-\varepsilon_{yy}$, $2\varepsilon_{xy}$) and frozen zone-center $E_{2g}$ phonons. Integrating out stable harmonic lattice modes generates a negative contribution to $v_M$ in \eqref{eq:FM_landau}, which could contribute to the stabilization of the AFM2 phase. We will return to this point in Section \ref{sec:magnetoelastic} where the magnitude of this coupling is estimated.

An off-diagonal bilinear $Q_{ab}$ carries a wave vector at the third $M$-point $M_c$ and transforms as the totally symmetric even small irrep $M^+_1$, allowing its coupling to frozen $M^+_1$ phonon modes at $M_c$. This coupling generates a negative contribution to $w_M$, which favors the collinear $3M$ phase. Sufficiently strong magnetoelastic coupling can also make the quartic part of \eqref{eq:FM_landau} non-positive-definite, leading to a first-order phase transition. A detailed analysis of these effects is outside the scope of the present study.

At $M_b$, the little space group $G_{M_b}$ retains an in-plane twofold axis $C_{2x}$ that leaves each Mn1 site invariant up to a lattice translation with unit phase factor at $M_b$, while interchanging the Mn2 sublattices $A$ and $C$. Since the corresponding AFM2 mode is odd under this operation [see \eqref{eq:Mb_pattern}], its exchange field on each Mn1 atom vanishes. The same statement applies to symmetry-related AFM2 modes at $M_a$ and $M_c$, and hence to any linear combination of the three modes. Thus, any superposition of AFM2 modes in the $M$ star has no bilinear coupling to the Mn1 ordering amplitudes, as anticipated above in Section \ref{subsec:GeneralSymmetry}.

The analysis of this section shows that the emergence of the seemingly exotic AFM2 phase in \MnSi\ is not accidental: its experimental observation tells us that the first eigenvalue of $\mathcal K(\q)$ to vanish on cooling corresponds to the symmetry-protected inversion-even, permutation-odd eigenmode at the $M$ star, while higher-order terms select the single-arm region of Fig. \ref{fig:MPD}. It is therefore unlikely that the sublattice that is forced by symmetry to remain disordered \emph{also} loses its fluctuating local moment for reasons unrelated to symmetry, as it is sometimes assumed in density-functional calculations \cite{dosSantos2021,MendozaEstrada2025}. This interpretation is consistent with the observation of coexisting spin waves and diffuse spin fluctuations in the AFM2 phase \cite{Biniskos2018,Biniskos2023}.

\subsection{Zone-center $E_{2g}$ ordering}
\label{subsec:Glandau}

To explain the observed AHE, it was proposed in Ref.\ \cite{Reichlova2024} that magnetic ordering in thin \MnSi\ films switches from the bulk AFM2 phase considered in Section \ref{subsec:Mlandau} to a zone-center pattern with the same $(1,-1,0)$ sublattice amplitudes. We now consider how this ordering could emerge from a Landau theory.

In the parity-even sector, the three ordering vectors $\m^+_A$, $\m^+_B$, $\m^+_C$ transform under the permutation representation $D^+(\Gamma)$ of the $6/mmm$ point group, with $D^+_g(\Gamma)\in S_3$, forming a tricolor point group \cite{Harker1981,Lifshitz-RMP}. Because, according to \eqref{eq:Kinvariance}, the nonrelativistic $\mathcal K(\Gamma)$ is invariant under this representation, the eigenspaces of $\mathcal K(\Gamma)$ correspond to the irreps of $6/mmm$ contained in $D^+(\Gamma)$. Those irreps are $A_{1g}$, which corresponds to the uniform magnetization $\m^+_A=\m^+_B=\m^+_C$, and a two-dimensional $E_{2g}$ irrep in the sublattice space whose eigenspace is defined by the constraint $\m^+_A+\m^+_B+\m^+_C=0$.

Because our target phase is compensated, the relevant irrep is $E_{2g}$. The corresponding order parameter space is described by two spin vectors. Choosing an orthonormal basis in the $E_{2g}$ subspace:
\begin{equation}
\e_1=\frac{1}{\sqrt{2}}(0,1,-1),\quad
\e_2=\frac{1}{\sqrt{6}}(2,-1,-1),
\end{equation}
we may write $(\m^+_A,\m^+_B,\m^+_C)=\e_1\otimes \Lop_1+\e_2\otimes\Lop_2$, or explicitly
\begin{align}
\m^+_A &= \sqrt{\frac{2}{3}}\,\Lop_2,\nonumber\\
\m^+_B &= \frac{1}{\sqrt{2}}\Lop_1-\frac{1}{\sqrt{6}}\Lop_2,\label{eq:Gamma_E_vectors}\\
\m^+_C &=-\frac{1}{\sqrt{2}}\Lop_1-\frac{1}{\sqrt{6}}\Lop_2.\nonumber
\end{align}

To quartic order, the Landau free energy is
\begin{equation}
F_\Gamma = a_\Gamma L^2 +u_\Gamma L^4
+v_\Gamma C_\times+
\cdots,
\label{eq:FGamma_quartic_vector}
\end{equation}
where $L^2=\Lop_1^2+\Lop_2^2$ and $C_\times=(\Lop_1\times\Lop_2)^2$. At $a_\Gamma<0$ (below $T_c$) the paramagnetic phase is unstable. However, because the quadratic term in $F_\Gamma$ is degenerate with respect to the direction of the order parameters in both spin space and $E_{2g}$ subspace, the magnetic ordering pattern depends on higher-order terms.
For $v_\Gamma>0$, the free energy minimum corresponds to collinear $\Lop_1$ and $\Lop_2$ (or one of them vanishing), while the direction in the $E_{2g}$ space remains arbitrary. For $v_\Gamma<0$, at fixed $L^2$ the $C_\times$ term is maximized in a noncollinear state with $\Lop_1\perp\Lop_2$ and $L_1=L_2$. According to \eqref{eq:Gamma_E_vectors}, this gives a 120-degree ``clock'' state with equal moment magnitudes on the three sublattices forming an equilateral triangle in spin space. 
However, this is not the right phase in the context of \MnSi.

In the collinear $v_\Gamma>0$ branch, the order parameter can be represented by an angle in the $E_{2g}$ subspace:
\begin{equation}
\Lop_1=L\cos\theta\,\nhat,\quad
\Lop_2=L\sin\theta\,\nhat,
\label{eq:Gamma_collinear}
\end{equation}
with an arbitrary common spin direction $\nhat$.
The Landau theory of this collinear sector is identical to that of the stacked triangular Ising antiferromagnet (STAFI) model of the CsCoCl$_3$ and CsCoBr$_3$ compounds
\cite{TodorokiMiyashita2004}.
Up to fourth order, the nonrelativistic free energy $F_\Gamma$ \eqref{eq:FGamma_quartic_vector} is degenerate in $\theta$, meaning that $(0,1,-1)$ and $(2,-1,-1)$ ordering patterns span a degenerate space. The leading $\theta$-dependent free-energy term is of sixth order:
\begin{equation}
\Delta F^\mathrm{coll}_\G=w_{6}L^6\cos 6\theta.
\end{equation}
The sign of $w_{6}$ selects between two sets of symmetry-related orientations in the $E_{2g}$ subspace. At $w_6<0$, one of the stable directions is $\theta=0$, which, according to Eq.~\eqref{eq:Gamma_E_vectors}, gives the AFM2-like ordering pattern $(0,1,-1)$. The other five symmetry-equivalent choices in the $\theta=n\pi/3$ set are obtained by permutations. The angular dependence of AHE in thin \MnSi\ films has been attributed to these six domains \cite{Rial2024variants}.
In contrast, at $w_6>0$ one of the stable directions is $\theta=\pi/2$, which gives the $(2,-1,-1)$  ordering pattern.

Thus, altermagnetic AFM2-like ordering with the $(0,1,-1)$ pattern can emerge from the generic Landau free energy \eqref{eq:FGamma_quartic_vector} for the $E_{2g}$ exchange instability as long as higher-order terms have the right signs, i.e., $v_\G>0$ to pick the collinear branch and $w_6<0$ to select the $(0,1,-1)$ direction within the $E_{2g}$ irrep space.

Such ordering on a tripartite lattice under a two-dimensional irrep differs from the standard Landau theory of collinear altermagnetism, in which the N\'eel vector transforms as a nontrivial real one-dimensional irrep of the paramagnetic point group \cite{McClarty2024,Schiff2025}, and yet it is clearly altermagnetic. The distinction is that the paramagnetic phase here has higher symmetry than the  altermagnetic phase in which the two sublattices have been made equivalent: the magnetic transition itself reduces the full tripartite permutation symmetry to a bipartite one.

As in the $M$-point case considered in Section \ref{subsec:Mlandau}, bilinear coupling of the zone-center AFM2-like mode to Mn1 ordered moments is forbidden by the in-plane twofold axis with respect to which that mode is odd.

The mean-field single-site entropy contributes $u^\mathrm{ent}_\Gamma=(9/20)k_BT$ and $v^\mathrm{ent}_\Gamma=-(3/5)k_BT<0$ per primitive cell to the quartic terms, which favors the equal-amplitude orthogonal state. Similar to the $M$-star case, a competing contribution can come from magnetoelastic coupling. Indeed, the pair $Q_1=\Lop_1^2-\Lop_2^2$, $Q_2=2\Lop_1\cdot\Lop_2$ transforms as an $E_{2g}$ doublet which, similar to the two independent components of $Q_\Delta$ in the $M$-star case, can couple to the $E_{2g}$ in-plane shear strain. This coupling results in a negative contribution $F_\Gamma \propto Q^2=(Q_1^2+Q_2^2)$ due to ordering striction. Because $Q^2=L^4-4C_\times$, it makes a positive contribution to $v_\Gamma$.

Various mechanisms contribute to $w_6$. Single-site entropy gives $w^\mathrm{ent}_6=-(11/700)k_BT<0$ per primitive cell, favoring the $(0,1,-1)$ pattern. A third-order term in $L$ can couple to the uniform magnetization mode, which in the collinear case \eqref{eq:Gamma_collinear} results in the free-energy term $\propto L^3\M\cdot\nhat\sin3\theta$. This term reflects weak ferrimagnetism of the $(2,-1,-1)$ mode with $M\propto L^3$. Eliminating $\M$ results in a positive term in $w_6$ favoring $(2,-1,-1)$. Magnetostructural coupling can also contribute to $w_6$.

\section{Paramagnetic instability in M\MakeLowercase{n}$_5$S\MakeLowercase{i}$_3$ from DFT}
\label{sec:dlm}

The previous section established that the experimental AFM2 phase corresponds to an ordering eigenmode  that is odd under a sublattice swap that is matched to the given arm of the $M$ star, while the AFM2-like altermagnetic phase without cell doubling corresponds to the zone-center $E_{2g}$ eigenmode, subject to certain restrictions on higher-order free energy terms.

Partially ordered phases present a challenge for first-principles calculations involving a comparison of the total energies of different ordered configurations. Since the AFM2 phase does not survive to zero temperature and is not fully ordered, this approach is not easily applicable to it. The key question is not which ordered structure has the lowest energy, but which partially ordered phase emerges from the hexagonal paramagnet under cooling. Because the AFM2 phase appears via a phase transition that is only weakly first-order, its ordering pattern is expected to reflect the leading quadratic instability of the paramagnetic phase.

In this section, the linear-response technique \cite{Staunton1986,LKAG,Ondracek2010} is used to calculate the exchange kernel $J_{\alpha\beta}(\q)$ in the paramagnetic DLM reference state \cite{OguchiTerakuraHamada1983,GyorffyPindorStauntonStocksWinter1985,Turek1997} treated within density functional theory. The identification of the resulting exchange instabilities provides a natural connection to the Landau theory.

\subsection{Computational details}

The electronic structure was described using the scalar-relativistic tight-binding linear muffin-tin orbital (TB-LMTO) method \cite{Andersen1975} in the atomic spheres approximation (ASA) along with the DLM technique \cite{Turek1997} implemented in the Questaal code \cite{Ke2013,Pashov2020:CPC}. We used the experimental structural parameters $a=6.910$ \AA, $c=4.814$ \AA, $x_\mathrm{Mn2}=0.2360$, and $x_\mathrm{Si}=0.5991$ \cite{Lander01091967}. The atomic sphere radii were chosen as 1.411, 1.420, and 1.435 \AA\ for Mn2, Mn1, and Si atoms, respectively, and an empty sphere with a radius of 0.907 \AA\ was placed at the $2b$ Wyckoff position at the center of each Mn2 octahedron. Basis states up to $l=2$ were included for all physical atoms and up to $l=1$ for the empty spheres. 

With the local density approximation \cite{vonBarthHedin1972}, the self-consistent DLM state has a $\mu_\mathrm{Mn2}=2.54 \mu_B$ local moment on the Mn2 sites, while the Mn1 atoms have no magnetic moment. With the generalized-gradient approximation (GGA) \cite{PerdewBurkeErnzerhof1996}, $\mu_\mathrm{Mn2}$ increases to 2.85 $\mu_B$.

Experimentally, the ordered moment on two thirds of the Mn2 sites in the AFM2 phase was estimated as 1.48 $\mu_B$ at 70 K \cite{BrownForsyth1995}, which is about 70\% of the AFM2 phase transition temperature. In the low-temperature AFM1 phase, the moments on the ordered Mn2 sites were estimated as 2.30 and 1.85 $\mu_B$ at 4 K \cite{BrownForsythNunezTasset1992}. If we assume Heisenberg-model-like behavior with fluctuating local moments depending weakly on temperature and magnetic configuration, these measurements suggest that $\mu_\mathrm{Mn2}$ is in the 1.9--2.3 $\mu_B$ range. Thus, the LDA local moment is much closer to experiment than the GGA value. To further test the sensitivity of $J_{\alpha\beta}(\q)$ to the magnitude of the local moment, additional calculations were performed with the spin-dependent part of the LDA exchange-correlation potential $B_{xc}$ scaled \cite{Ortenzi2012} by a factor $s_{xc}$ varying between 0.9 and 1. In particular, scaling by $s_{xc}=0.95$ reduces the self-consistent $\mu_\mathrm{Mn2}$ to 2.14 $\mu_B$.

The exchange parameters were calculated \cite{Pashov2020:CPC} using a $12\times12\times16$ Brillouin zone mesh and a 31-point elliptical quadrature for the energy integration, which provided excellent numerical convergence.

\subsection{Exchange interaction in the DLM state}

The exchange parameters are defined so that the magnetic energy is $E=-\sum_{ij} J_{ij} \hat\e_i\cdot\hat\e_j$, where the sum over Mn2 sites is unrestricted, and $\hat\e_i$ is a unit vector representing the orientation of the local moment on site $i$. Because $J_{ij}$ are translationally invariant, we can relabel them as $J_{\alpha\beta}(\T_{ij})$, where $\T_{ij}$ is the translation from the primitive cell containing site $i\in\alpha$ to the one containing $j\in\beta$.

Table \ref{tab:mc_truncated_jij} lists the exchange parameters exceeding 0.02 mRy in magnitude calculated in the DLM state of \MnSi\ using the LDA potential; the four nearest neighbors are illustrated in Fig. \ref{fig:Jsketch}. As expected, the strongest coupling is the geometrically frustrated antiferromagnetic exchange $J_1\approx-0.89$ mRy within the nearest-neighbor $ABC$ triangles. The second largest is the ferromagnetic $J_3\approx0.21$ mRy between $X$ and $\bar X$ in the same primitive cell, which favors the parity-even sector at both $\Gamma$ and $M$ points. We also find a fairly strong \emph{antiferromagnetic} coupling $J_4\approx-0.16$ mRy for $X$--$\bar X$ pairs in the \emph{neighboring} octahedral chains. Within the parity-even sector, strong antiferromagnetic $J_4$ implies that $\Gamma$-point ordering should be substantially less favorable compared to zone-boundary ordering. 

\begin{table}[htb]
\centering
\small
\setlength{\tabcolsep}{4pt}
\renewcommand{\arraystretch}{1.08}
\caption{Exchange parameters $J_N$ in the DLM state, where $N$ is the bond-star index ordered by nondecreasing distance, with $|J_N|>0.02$ mRy, calculated with the LDA potential and labeled as $J_{\alpha\beta}(\T_{ij})$ (see text). Each row lists a representative pair as a combination of $\alpha$--$\beta$ and $\T_{ij}$. $R_{ij}$ is the bond length, and $z$ is the coordination number.
Note that $\mathbf T=-\mathbf a_1$ connects the central cell to the neighboring cell in Fig. \ref{fig:Jsketch}.}
\label{tab:mc_truncated_jij}
\begin{tabular}{cccccc}
\hline\hline
&
$\alpha$--$\beta$
& $\T_{ij}$
& $R_{ij}$ (\AA)
& $z$
& $J_{ij}$ (mRy)
\\
\hline
$J_1$
& $A$--$B$
& $\mathbf 0$
& $2.82$
& $2$
& $-0.887$
\\
$J_2$
& $A$--$\bar B$
& $\mathbf 0$
& $2.91$
& $4$
& $+0.055$
\\
$J_3$
& $A$--$\bar A$
& $\mathbf 0$
& $4.05$
& $2$
& $+0.210$
\\
$J_4$
& $A$--$\bar A$
& $-\mathbf a_1$
& $4.37$
& $2$
& $-0.158$
\\
$J_5$
& $A$--$B$
& $-\mathbf a_1$
& $4.68$
& $4$
& $+0.044$
\\
$J_6$
& $A$--$A$
& $\mathbf a_3$
& $4.81$
& $2$
& $-0.020$
\\
$J_7$
& $A$--$B$
& $\mathbf a_3$
& $5.58$
& $4$
& $-0.058$
\\
$J_8$
& $\bar B$--$C$
& $-\mathbf a_1$
& $5.80$
& $4$
& $-0.056$
\\
$J_{13}$
& $B$--$B$
& $-\mathbf a_1$
& $6.91$
& $4$
& $-0.036$
\\
$J_{24}$
& $A$--$C$
& $-(\mathbf a_1+\mathbf a_2)$
& $9.14$
& $2$
& $+0.073$\\
\hline\hline
\end{tabular}
\end{table}

The DLM exchange parameters listed in Table \ref{tab:mc_truncated_jij} may be compared with calculations for the symmetry-broken orthorhombic reference state corresponding to the fully ordered AFM2 phase in which the disordered Mn2 sites are assumed to have vanishing local moments \cite{dosSantos2021}. 
Among the first four $J_N$ parameters reported in Ref.~\cite{dosSantos2021}, $J_1\approx-0.90$ mRy, $J_3\approx0.29$ mRy, and $J_4\approx-0.21$ mRy (the latter for the antiparallel spin pair) have the same signs and are quantitatively similar to the corresponding DLM values in Table \ref{tab:mc_truncated_jij}, despite the drastic difference in the reference state. The comparison is particularly notable because one of the two $J_1$ neighbors is nonmagnetic in the ordered AFM2 reference state. Their $J_2\approx-0.16$ mRy, which likewise connects only two rather than four magnetic neighbors, instead has the opposite sign. We emphasize that the two phases have different magnetic networks and different symmetries; only the DLM exchange parameters respecting full paramagnetic symmetry can be used to characterize the magnetic phase transition. 

\subsection{Ordering eigenmodes}

The Fourier-transformed exchange matrix is defined as
\begin{equation}
J_{\alpha\beta}(\q)=\sum_{\T}J_{\alpha\beta}(\T)
e^{i\q\T}.
\end{equation}
Its largest eigenvalue $\lambda_{\max}$ determines the leading mean-field instability of the paramagnetic phase in the classical Heisenberg model. The corresponding mean-field critical temperature of the second-order phase transition is $k_BT_c^{\rm MF}=\frac{2}{3}\lambda_{\max}$.

\begin{figure}[htb]
\centering
\includegraphics[width=0.85\linewidth]{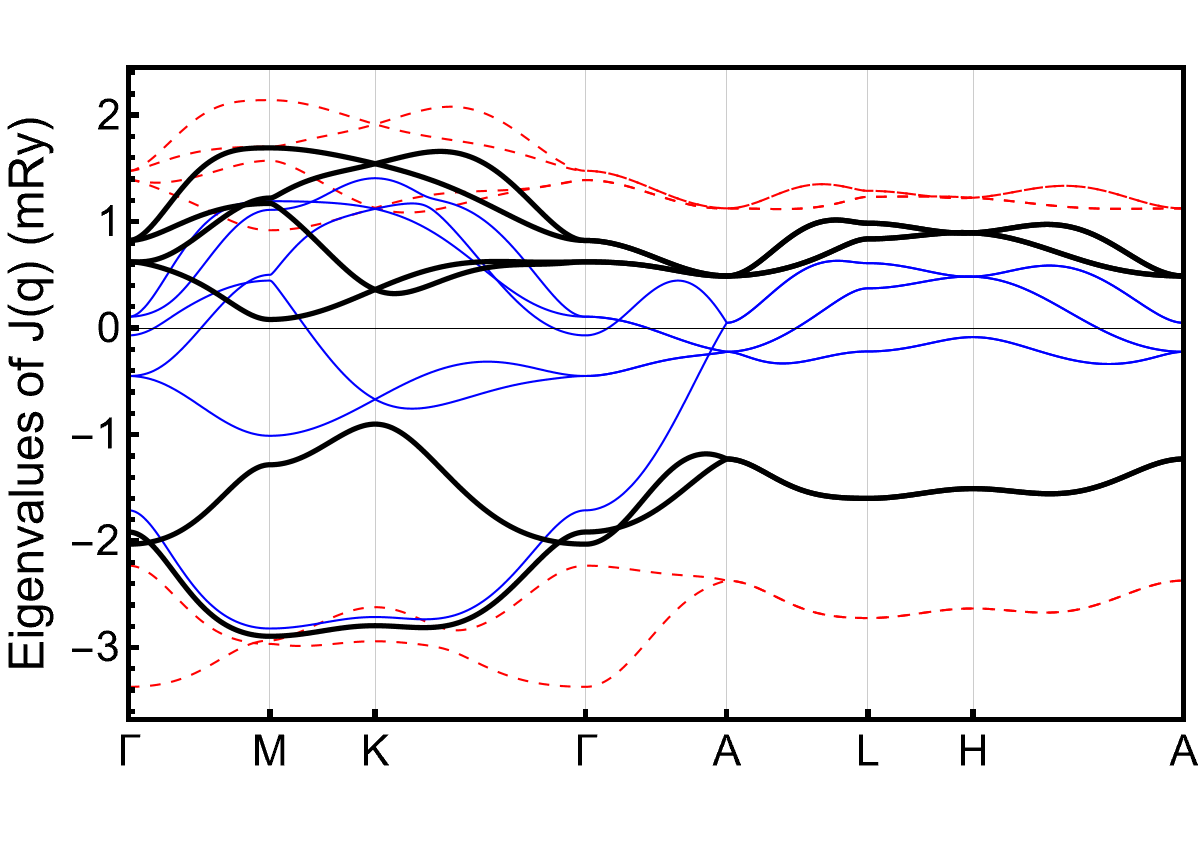}
\caption{Eigenvalues $\lambda_i(\q)$ of the six-sublattice $J_{\alpha\beta}(\q)$ exchange matrix calculated using the DLM method for unstrained Mn$_5$Si$_3$ along a high-symmetry path in the Brillouin zone. Bold black lines: LDA, with $\mu_\mathrm{Mn2}=2.54 \mu_B$. Dashed red lines: LDA with $B_{xc}$ scaled by $s_{xc}=0.95$, leading to $\mu_\mathrm{Mn2}=2.14 \mu_B$. Solid blue lines: GGA, with $\mu_\mathrm{Mn2}=2.85 \mu_B$. In the former two cases, the maximum eigenvalue occurs at $M$, and the corresponding eigenvector is parity-even with the AFM2 pattern $(0,1,-1)$. For GGA, the maximum eigenvalue is at $K$, and the corresponding eigenvector is nearly uniform within the primitive cell.
}
\label{fig:Jq_unstrained}
\end{figure}

Figure~\ref{fig:Jq_unstrained} shows the eigenvalues $\lambda_i(\q)$ of $J_{\alpha\beta}(\q)$ along high-symmetry directions for the three cases: the LDA potential ($\mu_\mathrm{Mn2}=2.54 \mu_B$), the scaled LDA potential ($\mu_\mathrm{Mn2}=2.14 \mu_B$), and GGA ($\mu_\mathrm{Mn2}=2.85 \mu_B$). In the former two cases, the largest eigenvalue occurs at the $M$ point; the corresponding eigenvector is parity-even under inversion and has the symmetry-protected $(1,-1,0)$ sublattice structure which, as explained in Section \ref{subsec:Mlandau}, leads to the experimental AFM2 phase if only a single $M$ arm condenses.

For the GGA potential, the largest eigenvalue is at $K$, which comes from the mode that is nearly ferromagnetic within a primitive cell and would produce, by a superposition with a degenerate mode at $\q_{K'}=-\q_K$, a noncollinear ordering with a $\sqrt3\times\sqrt3$ supercell. The same branch of the eigenvalue spectrum, with a maximum at $K$, is visible for the LDA potential where it is stable at all $\q$, and the destabilization of this branch is the most pronounced effect of the local moment increase from 2.54 to 2.85 $\mu_B$. However, given that the GGA local moment is too large compared to experiment, it appears unlikely that the $K$-point mode could be relevant in bulk \MnSi.

For all three potentials, the largest eigenvalue at the $\Gamma$ point corresponds to the $E_{2g}$ doublet considered in Section \ref{subsec:Glandau}. In both conventional ($s_{xc}=1$) and scaled ($s_{xc}=0.95$) LDA ($\mu_\mathrm{Mn2}=2.54$ or 2.14 $\mu_B$), this $E_{2g}$ mode lies considerably below the leading eigenmode at $M$. In GGA, this mode is only slightly unstable, lying far below zone-boundary modes on the $KM$ line. In all three cases, the largest eigenvalue $\lambda_{\max}(\q)$ has a global \emph{minimum} at $\Gamma$ in the whole $q_z=0$ plane.

\begin{figure}[htb]
\centering
\includegraphics[width=0.84\linewidth]{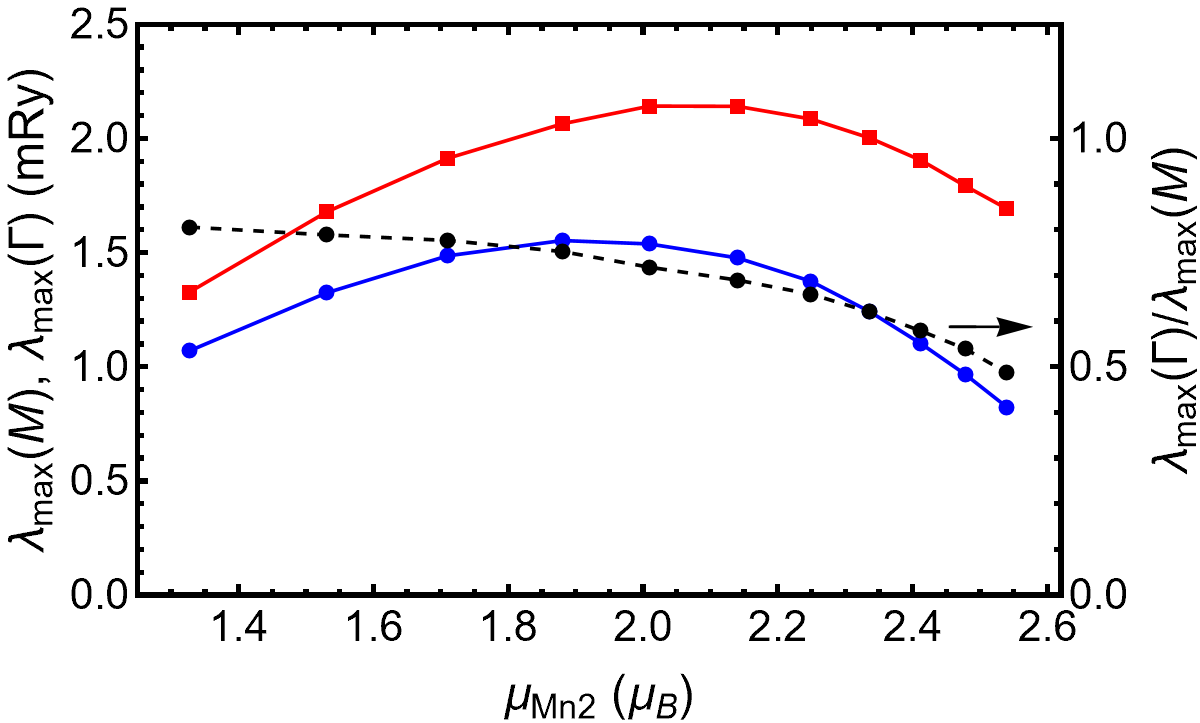}
\caption{The largest eigenvalues $\lambda_\mathrm{max}(M)$ (red line and squares) and $\lambda_\mathrm{max}(\Gamma)$ (blue line and circles), and their ratio $\lambda_\mathrm{max}(\Gamma)/\lambda_\mathrm{max}(M)$ (black dashed line and circles; right axis) as a function of the local magnetic moment $\mu_\mathrm{Mn2}$.
}
\label{fig:lambdaratio}
\end{figure}

Figure \ref{fig:lambdaratio} shows the largest eigenvalues $\lambda_\mathrm{max}(M)$ and $\lambda_\mathrm{max}(\Gamma)$, along with their ratio, as a function of $\mu_\mathrm{Mn2}$, which were calculated by varying the $B_{xc}$ scaling factor $s_{xc}$ between 0.9 and 1 in increments of 0.01. Although the $\lambda_\mathrm{max}(\Gamma)/\lambda_\mathrm{max}(M)$ ratio increases with decreasing $\mu_\mathrm{Mn2}$, it does not exceed 0.81 in the entire range of $\mu_\mathrm{Mn2}$ shown in Fig. \ref{fig:lambdaratio} and is less than 0.75 in the experimentally relevant range $\mu_\mathrm{Mn2}\gtrsim1.9 \mu_B$. Thus, in the DLM picture of the paramagnetic state, the $E_{2g}$ instability at $\Gamma$ is not competitive with the $M$-point instability in bulk \MnSi.

For the LDA calculation, the origin of the $M$-point preference can be seen directly from  $\lambda_\mathrm{max}(M)-\lambda_\mathrm{max}(\Gamma)=4(-J_4+J_5-J_{13})+\cdots$. All other parameters listed in Table \ref{tab:mc_truncated_jij} cancel in the difference. With the signs from Table \ref{tab:mc_truncated_jij}, each of the three displayed terms favors ordering at the $M$ point compared to $\Gamma$.

\subsection{Effect of epitaxial strain}

Thin \MnSi\ films grown by molecular beam epitaxy on Si(111) exhibit epitaxial tensile strain on the order of 1\% due to the thermal expansion mismatch, with the accompanying contraction of the $c$ axis by 0.5--0.6\% \cite{Reichlova2024}. The films grown by magnetron sputtering on Al$_2$O$_3$  \cite{Han2025PRA} had lattice constants $a$ and $c$ that are both $\sim0.3$\% larger than those reported in Ref. \cite{Reichlova2024}, suggesting some degree of disorder or off-stoichiometry. 

\begin{figure}
    \centering
    \includegraphics[width=0.85\linewidth]{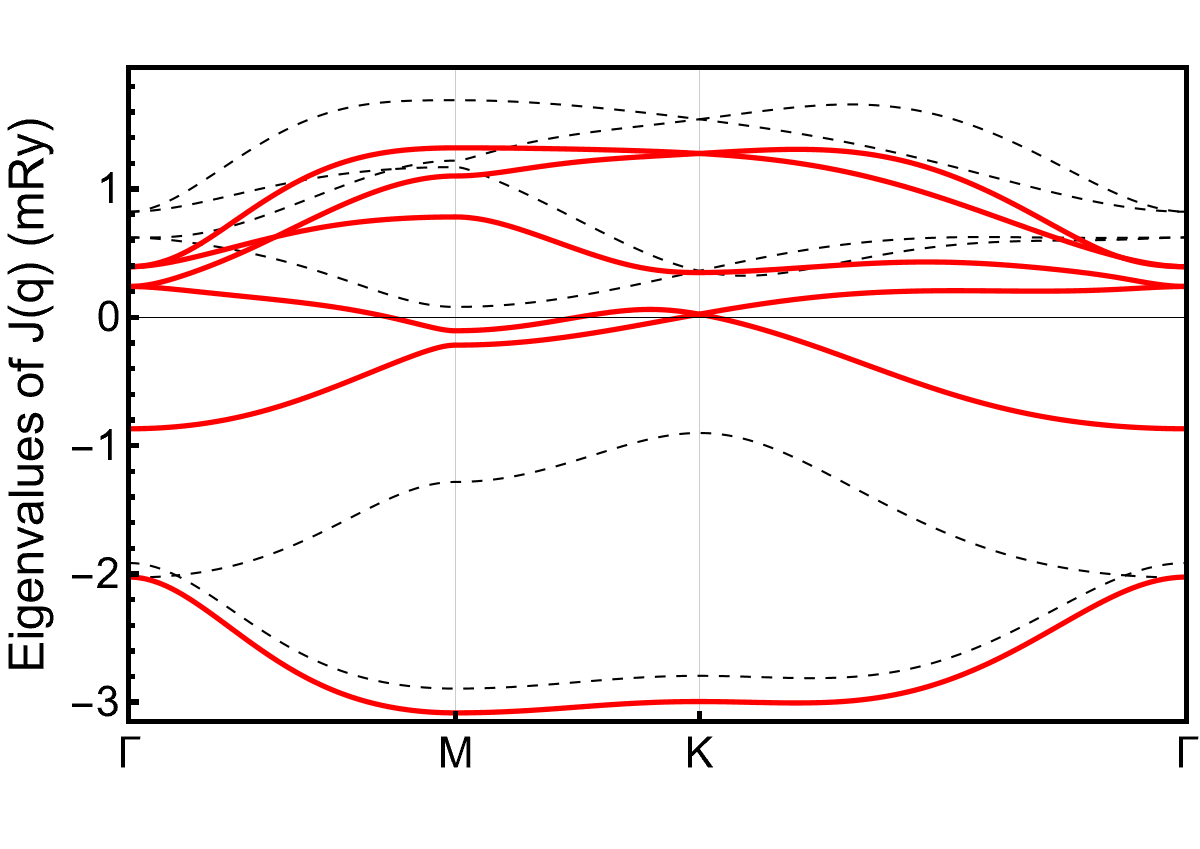}
    \caption{Eigenvalues of $J_{\alpha\beta}(\q)$ along the $\Gamma$KM$\Gamma$ path from LDA. Dashed black lines: bulk lattice constants (same as solid black lines in Fig. \ref{fig:Jq_unstrained}). Solid red lines: 1\% in-plane expansion combined with $-0.3$\% out-of-plane contraction.}
    \label{fig:strainedJq}
\end{figure}

The elastic constants calculated for the ferromagnetic state \cite{deJong2015Elastic,deJong2016ElasticData} suggest an out-of-plane to in-plane strain ratio close to $-0.3$. Figure \ref{fig:strainedJq} shows the effect of a 1\% biaxial in-plane expansion combined with a $0.3$\% $c$-axis contraction on the eigenvalues of $J_{\alpha\beta}(\q)$ for the LDA potential; the fractional coordinates were kept fixed because the ASA-DLM technique does not provide reliable forces. It was checked that doubling the $c$-axis contraction to $0.6$\% results in a very similar spectrum. We see that epitaxial strain of this magnitude lowers the upper $\lambda_i(\q)$ branches but does not qualitatively change their dispersions, except that the largest eigenvalues at the $M$ and $K$ points become closer to each other, and $\lambda_{\max}(\q)$ becomes flatter near the zone boundary at $q_z=0$. The $E_{2g}$ mode at $\Gamma$ becomes even less competitive compared to the unstrained lattice. These trends are likely mediated, at least in part, by the enhancement of the local moment $\mu_\mathrm{Mn2}$ from 2.54 to 2.68 $\mu_B$ by the strain; indeed, Fig. \ref{fig:lambdaratio} shows that $\lambda_\mathrm{max}(M)$, $\lambda_\mathrm{max}(\G)$, and the $\lambda_\mathrm{max}(\G)/\lambda_\mathrm{max}(M)$ ratio all decline with increasing $\mu_\mathrm{Mn2}$ in this range. A similar trend was found for the scaled LDA potential with $s_{xc}=0.95$: the same strain increases $\mu_\mathrm{Mn2}$  from 2.14 to 2.39 $\mu_B$, decreases $\lambda_\mathrm{max}(M)$ from 2.14 to 1.86 mRy, and $\lambda_\mathrm{max}(\G)$ from 1.48 to 1.16 mRy.

Thus, the DLM picture does not support the stabilization of the AFM2-like altermagnetic phase by tensile epitaxial strain.
Because $\lambda_{\max}(\q)$ is notably flattened at the periphery of the $q_z=0$ cross-section of the Brillouin zone under in-plane lattice expansion, it is conceivable that higher-order or entropic effects could stabilize a different phase. However, there appears to be no mechanism in the strain-dependent $J(\q)$ to sustain an AHE-active magnetic state to more than 200 K.

\section{Monte Carlo simulations and estimated ordering temperature}
\label{sec:MC}

The leading $M$-point eigenvalue is approximately
$\lambda_{\max}(M)\approx 1.69~{\rm mRy}$ with the LDA potential or 2.14 mRy with $s_{xc}=0.95$. The corresponding range of $T_c^\mathrm{MF}$ is 180--225 K. Given the small coordination number and  strong frustration of the nearest-neighbor coupling, it should be expected that the mean-field theory strongly overestimates the transition temperature. 

To estimate it more accurately, we performed classical Monte Carlo simulations using the UppASD code \cite{UppASD} with the LDA exchange parameters. Only exchange parameters with $|J_{ij}|>0.02$ mRy, which are listed in Table \ref{tab:mc_truncated_jij}, were included; this truncation only slightly increases $\lambda_{\max}(M)$ by about 1\%.
We used periodic $12\times12\times8$ and $16\times16\times12$ supercells, containing 6912 and 18432 classical spins, respectively. At each temperature, the spins were initialized in random directions and equilibrated using the heat-bath algorithm for $2\times10^4$ Monte Carlo sweeps, followed by $5\times10^4$ measurement sweeps. Spin configurations were saved every 500 sweeps, yielding 100 configurations per run for analysis. Vector order parameters $\Lop_\nu$ were obtained by projecting these configurations onto the AFM2 modes at the three $M_\nu$ points.

\begin{figure}[htb]
    \centering
    \includegraphics[width=0.85\linewidth]{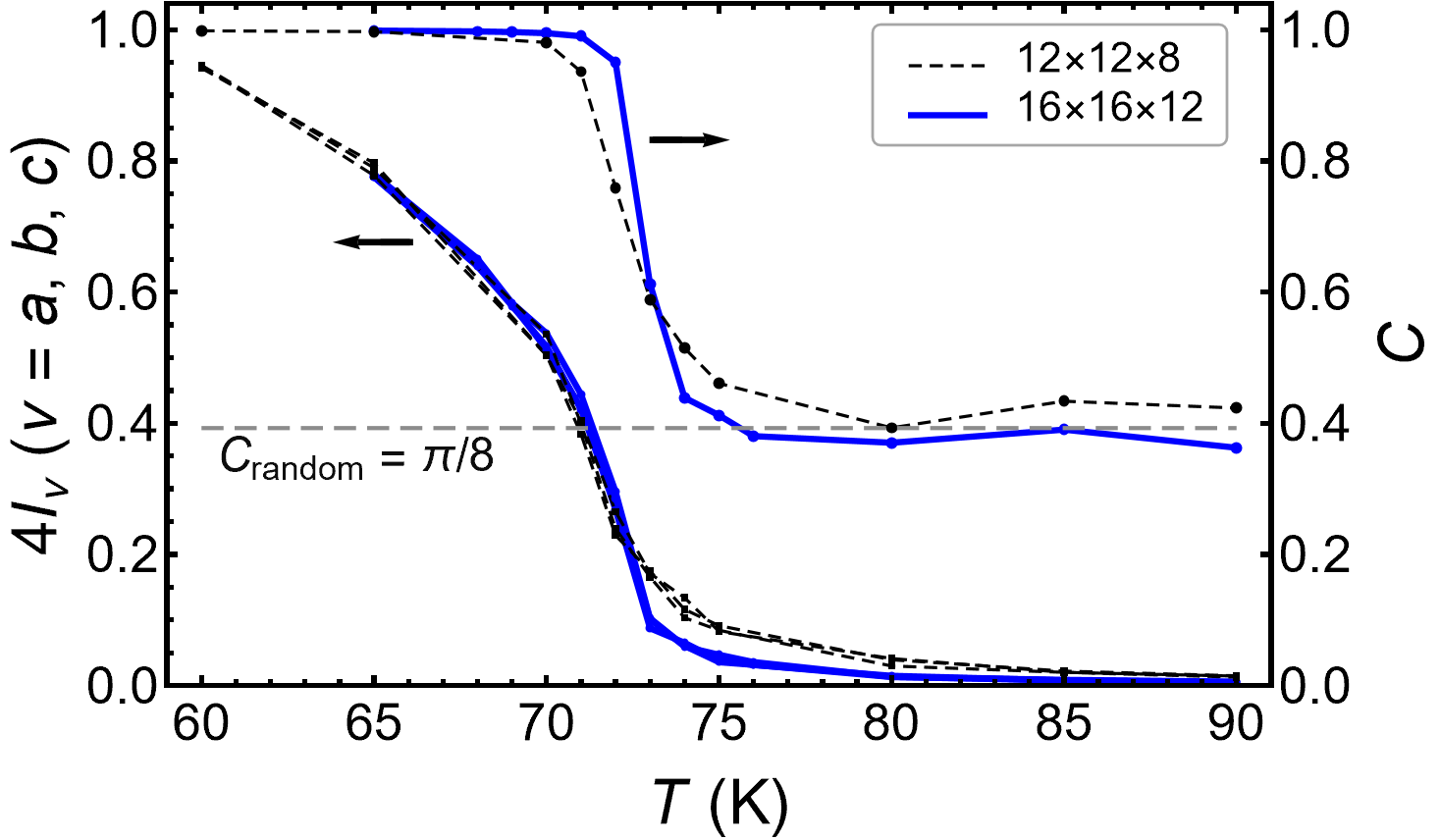}
    \caption{Temperature dependence of the arm-resolved intensities $I_\nu=\langle\Lop_\nu^2\rangle$ (left axis; scaled by a factor of 4 for visual clarity) and of the normalized absolute $M$-arm chirality $C=\langle|\Lop_a\cdot(\Lop_b\times\Lop_c)|\rangle/\langle L_aL_bL_c\rangle$ (right axis), obtained from classical Monte Carlo simulations. Dashed black and solid blue curves correspond to the $12\times12\times8$ and $16\times16\times12$ supercells, respectively. The horizontal dashed line marks the random-orientation expectation $C_{\rm random}=\pi/8$. The three $I_\nu$ curves for a given supercell size are nearly identical.}
    \label{fig:MC}
\end{figure}

Figure \ref{fig:MC} shows the three arm-resolved intensities $I_\nu=\langle\Lop_\nu^2\rangle$ and the normalized absolute $M$-arm chirality $C=\langle|\Lop_a\cdot[\Lop_b\times\Lop_c]|\rangle/\langle L_aL_bL_c \rangle$ averaged over 100 saved configurations, as a function of temperature. The three arm intensities are nearly equal and decrease rapidly near the phase transition, which we estimate at $T_N\approx73$ K. The value $C\approx1$ in the ordered phase indicates that the $M$-arm order parameters are nearly orthogonal; above the phase transition it converges to the value $C_\mathrm{random}=\pi/8$ corresponding to a random ensemble of vector triplets. This behavior identifies the ordered phase as the equal-amplitude orthogonal $3M$ ordering. An inspection of the ordered configurations confirmed this assignment.

The emergence of the orthogonal $3M$ phase in the classical  Heisenberg model is natural, because, as explained above in Section \ref{subsec:Mlandau}, single-site entropy produces quartic free-energy terms in the positive quadrant of the phase diagram of Fig. \ref{fig:MPD}; the single-arm AFM2 state requires additional quartic interactions beyond that model.
However, it is reasonable to expect that the transition temperature would be of the same order regardless of which phase is selected by quartic terms. Therefore, $T_N\approx73$ K predicted by Monte Carlo simulations is in reasonable agreement with the experimental $T_{N2}\approx100$ K. The 27\% larger $\lambda_{\max}(M)$ obtained with $s_{xc}=0.95$, together with the similar overall dispersion of the eigenvalue spectrum, is expected to increase the Monte Carlo $T_N$ and further improve agreement with experiment. Thus, the DLM state with LDA or scaled LDA ($s_{xc}=0.95$) potentials yields a reasonable range of local moments, an exchange kernel that correctly reproduces the dominant bulk $M$-point instability, and a quantitatively reasonable ordering temperature. 

\section{Magnetoelastic coupling}
\label{sec:magnetoelastic}

As explained in Section \ref{subsec:Mlandau}, magnetostructural coupling can contribute to the effective quartic parameters in the Landau free energy \eqref{eq:FM_landau}, and, in particular, coupling to $E_{2g}$ in-plane shear strain components can add a negative contribution to $v_M$, pushing the system toward the experimentally observed single-arm AFM2 phase (see Fig. \ref{fig:MPD}). In this section, we provide a semiquantitative estimate of this coupling.

Let $q_\nu=|\mathbf L_\nu|^2$, where $\Lop_\nu$ is normalized to the local moment on the Mn2 atoms, and introduce a normalized $E_{2g}$ doublet $\Phi_1=(2q_b-q_a-q_c)/\sqrt6$, $\Phi_2=(q_a-q_c)/\sqrt2$ and the corresponding $E_{2g}$ strain components $\varepsilon_1=\varepsilon_{xx}-\varepsilon_{yy}$, $\varepsilon_2=2\varepsilon_{xy}$.
The elastic and magnetoelastic energies per primitive cell can be written as
\begin{equation}
    F_{\mathrm{el}}
    =\frac12 C_{66}V\boldsymbol\varepsilon^2,
    \quad F_{\mathrm{me}}
    =\gamma\boldsymbol\varepsilon\cdot\boldsymbol\Phi
\label{Fme}
\end{equation}
where $V$ is the primitive-cell volume, $C_{66}=(C_{11}-C_{12})/2$ is the basal shear modulus, and $\gamma$ is the magnetoelastic coupling constant; $\boldsymbol\varepsilon=(\varepsilon_1,\varepsilon_2)$ and $\boldsymbol\Phi=(\Phi_1,\Phi_2)$ are vectors in the two-component $E_{2g}$ irrep space. Because $\Phi_1^2+\Phi_2^2=\mathrm{Tr}\,Q_\Delta^2$, eliminating the strain results in a magnetoelastic contribution to $v_M$ in \eqref{eq:FM_landau}: 
\begin{equation}
    \Delta v_M^{\mathrm{me}}
    =-\frac{\gamma^2}{2C_{66}V}.
    \label{eq:dvM_ME}
\end{equation}

The coupling $\gamma$ may be estimated from the strain-induced splitting of the exchange eigenvalues at the three arms of the $M$ star. To this end, we recalculated the largest eigenvalues $\lambda(M_\nu)$ of the exchange kernel $J(M_\nu)$ after imposing a volume-preserving orthorhombic strain $\varepsilon_1\equiv\varepsilon_{xx}-\varepsilon_{yy}=\pm0.01$. Because the internal relaxations cannot be easily calculated for the paramagnetic phase, the  fractional atomic coordinates are kept fixed.
Such strain leaves $M_a$ and $M_c$ equivalent, and we extract the linear slope of $\Delta\lambda\equiv\lambda(M_b)-\lambda(M_{a,c})$ with respect to $\varepsilon_1$.
With the normalization of Eqs. \eqref{eq:Mstar_realspace} and our definition of $J_{ij}$, the quadratic exchange energy per primitive cell is
$F_{\mathrm{ex}}=-4\sum_{\nu}\lambda(M_\nu)q_\nu$, and the part attributable to the $E_{2g}$ strain is $\Delta F_{\mathrm{ex}}=-(4\sqrt6/3)\Delta\lambda\,\Phi_1$.
Comparison with $F_{\mathrm{me}}$ in Eq. \eqref{Fme} therefore gives
\begin{equation}
    \gamma_\mathrm{cl}
    =-\frac{4\sqrt6}{3}
      \frac{d\Delta\lambda}{d\varepsilon_1}.
    \label{eq:gamma_from_splitting}
\end{equation}
where the subscript indicates that $\gamma_\mathrm{cl}$ is the clamped-ion estimate.
We found $\gamma_\mathrm{cl}\approx0.27$ eV per primitive cell in LDA or 0.29 eV with $s_{xc}=0.95$.

The equilibrium strain is $\boldsymbol\varepsilon=-\frac{\gamma}{C_{66}V}\boldsymbol\Phi$. In the single-$M$ AFM2 phase, it represents orthorhombic exchange striction. Its physical origin is that the antiferromagnetic exchange coupling within the nearest-neighbor $ABC$ triangles increases in magnitude when the bonds shorten. As a result, AFM2 ordering tends to contract the only kind of bond that connects the two magnetically ordered Mn2 sublattices.

For the AFM2 phase with the ordered moment $L_b$, we have
$\Phi_1=\sqrt{2/3}\,L_b^2$ and $\Phi_2=0$, and the exchange striction is
\begin{equation}
    \varepsilon_1
    =-\frac{\gamma}{C_{66}V}
      \sqrt{\frac23}\,L_b^2.
    \label{eq:spontaneous_strain}
\end{equation}
A single-crystal neutron diffraction study reported the exchange striction as a function of temperature, which reaches a magnitude of 0.481\% at 70 K, right above the phase transition into the AFM1 phase \cite{BrownForsyth1995}.

The elastic tensor of \MnSi\ was calculated for the ferromagnetic state in Refs. \cite{deJong2015Elastic,deJong2016ElasticData} and yielded a relatively soft $C_{66}\approx16.8$ GPa. Taking this value along with our estimated $\gamma_\mathrm{cl}$ and the experimental $|\varepsilon_1|\approx0.0048$ \cite{BrownForsyth1995}, we can use \eqref{eq:spontaneous_strain} to estimate $L_b\sim0.7$, which is reasonable at $T/T_N\approx0.7$ and roughly consistent with the measured ordered moment of 1.48 $\mu_B$ at 70 K \cite{BrownForsyth1995}. Thus, the above $C_{66}$ value, the calculated $\gamma_\mathrm{cl}$, and the experimental exchange striction are reasonably consistent among themselves. 

Using \eqref{eq:dvM_ME}, we then estimate $\Delta v_M^{\mathrm{me}}\simeq-1.8\ \mathrm{meV}$ per primitive cell. This should be compared with the classical single-site entropy contribution $v_M^{\mathrm{ent}}=\frac9{10}k_BT$, which is about 8 meV near $T_N\approx100$ K. Thus, magnetoelastic coupling of the unstable $M$-star modes to $E_{2g}$ strain provides a substantial negative contribution to $v_M$, which is roughly a quarter of the single-site entropy contribution.

To understand the role of the neglected internal relaxations, consider the magnetostructural  free energy depending explicitly on the amplitudes  $\mathbf{u}_\alpha$ of the five $E_{2g}$ phonon mode pairs, labeled by Greek indices:
\begin{align}
    F_\mathrm{ms}&=\frac12 C^\mathrm{cl}_{66}V\boldsymbol\varepsilon^2+\frac12\sum_{\alpha\beta}K_{\alpha\beta}\mathbf{u}_\alpha\cdot\mathbf{u}_\beta+\sum_\alpha  f_\alpha\boldsymbol\varepsilon\cdot\mathbf{u}_\alpha\nonumber\\
    &+\gamma_\mathrm{cl}\boldsymbol\varepsilon\cdot\boldsymbol\Phi+\sum_\alpha g_\alpha\mathbf{u}_\alpha\cdot\boldsymbol\Phi
\end{align}
where $C_{66}^\mathrm{cl}$ is the clamped-ion shear modulus, $K_{\alpha\beta}$ is the positive-definite effective force-constant matrix, $f_\alpha$ are the  strain-induced effective forces, and $g_\alpha$ are the effective Kanzaki forces coupling magnetic order to internal displacements. Minimizing over $\boldsymbol\varepsilon$ and $\mathbf{u}_\alpha$ and matching with \eqref{eq:FM_landau} gives
\begin{equation}
    \Delta v_M=-\frac{\gamma^2}{2C_{66}V}-\frac12\sum_{\alpha\beta}g_\alpha K^{-1}_{\alpha\beta}g_\beta
    \label{eq:dvm1}
\end{equation}
where $\gamma=\gamma_\mathrm{cl}-\sum_{\alpha\beta}f_\alpha  K^{-1}_{\alpha\beta}g_\beta$ and $C_{66}V=C^\mathrm{cl}_{66}V-\sum_{\alpha\beta}f_\alpha K^{-1}_{\alpha\beta}f_\beta$. The first term in \eqref{eq:dvm1} is $\Delta v_M^\mathrm{me}$ from \eqref{eq:dvM_ME}, and the second comes from direct coupling to $E_{2g}$ phonons.

Using the equilibrium exchange striction coefficient $R=\boldsymbol\varepsilon\cdot\boldsymbol\Phi/\boldsymbol\Phi^2=-\gamma/(C_{66}V)$, Eq. \eqref{eq:dvm1} may be written:
\begin{equation}
    \Delta v_M=-\frac12 C_{66}VR^2+\Delta v_M^\mathrm{int},
    \label{eq:dvm2}
\end{equation}
where $\Delta v_M^\mathrm{int}$ is the second term in \eqref{eq:dvm1}. 
Our earlier estimate of the first term $\Delta v_M^\mathrm{me}$ is roughly consistent with taking the striction $R$ inferred from experiment and $C_{66}$ from the calculation in the ferromagnetic state \cite{deJong2015Elastic,deJong2016ElasticData}; this latter interpretation is independent of the calculated $\gamma_\mathrm{cl}$. (Note that, at fixed $R$ inferred from experiment, our estimate of $|\Delta v_M^{\rm me}|$ is suppressed, rather than enhanced, by the anomalously soft calculated $C_{66}$.) At the same time, the full $\Delta v_M$ in \eqref{eq:dvm2} includes an \emph{additional} negative-definite term $\Delta v_M^\mathrm{int}$ coming from direct coupling to internal $E_{2g}$ displacements. Thus, the neglect of internal relaxations likely results in an underestimation of $|\Delta v_M|$.

To conclude this section, let us note that the exchange striction may be suppressed in a thin \MnSi\ film that is epitaxially clamped to a symmetry-preserving substrate, which would suppress the first term in \eqref{eq:dvm2}. Thus, the total $v_M$ coefficient may be larger in such a film compared to the bulk. In a fine-tuned parameter regime, the absence of the magnetoelastic term could even change the sign of $v_M$ from negative to positive, pushing the system across the phase boundary from the bulk single-arm AFM2 phase into the orthogonal $3M$ region in the phase diagram of Fig. \ref{fig:MPD}. Note that $\Phi_1=\Phi_2=0$ in the orthogonal $3M$ phase, and therefore, in the nonrelativistic limit considered here, it should exhibit no orthorhombic exchange striction.

\section{Discussion}
\label{sec:discussion}

DLM calculations with either LDA or a slightly downscaled exchange-correlation field give reasonable Mn2 local moments and reproduce the bulk $M$-point AFM2 instability.
The magnetic ordering temperature estimated from classical Monte Carlo simulations is also of the correct order of magnitude. Although these simulations result in the orthogonal $3M$ phase instead of AFM2, this does not contradict the quadratic instability analysis, because both phases condense from the same $M$-star exchange mode and are distinguished by quartic terms that are severely restricted by the pair-exchange model.

The DLM exchange kernel does not support altermagnetic $\Gamma$-point ordering with an AFM2-like $(1,-1,0)$ intra-cell pattern that was proposed to explain the AHE observed in thin \MnSi\ films \cite{Reichlova2024}. Although the largest eigenvalue $\lambda_{\max}(\G)$ of the exchange kernel $J_{\alpha\beta}(\G)$ belongs to the $E_{2g}$ representation, which, as shown in Section \ref{subsec:Glandau}, \emph{can} produce that altermagnetic phase, $\lambda_{\max}(\G)$ itself is significantly less than $\lambda_{\max}(M)$ and is a global \emph{minimum} within the $q_z=0$ plane. This property is robust against reasonable changes in the local moment and with respect to an in-plane lattice expansion corresponding to epitaxial \MnSi\ films grown on Si(111); in fact, such strain makes $\Gamma$-point ordering even \emph{less} favorable compared to the bulk.
DLM calculations also give no indication that tensile epitaxial strain generates a strongly enhanced magnetic exchange scale; instead, they point in the opposite direction: a 1\% tensile epitaxial strain typical for \MnSi\ films grown on Si(111) reduces the leading $M$-point exchange eigenvalue by about 20\%.

The spontaneous AHE in thin films establishes a time-reversal-breaking, Hall-allowed magnetic state, but it does not uniquely determine its microscopic magnetic structure or propagation vector. The angular response is consistent with the reported set of magnetic variants, but their microscopic realization remains uncertain.

Epitaxial clamping can in principle alter the selection among single-$M$, multi-$M$, or other nearby finite-$\q$ states by modifying quartic magnetoelastic terms and by exploiting the strain-induced flattening of $\lambda_{\max}(\q)$ near the zone boundary.
However, such rearrangements would inherit essentially the same quadratic exchange scale and therefore cannot naturally account for anomalous transport persisting to temperatures more than twice the bulk AFM2 transition.
It is noteworthy that Ref. \cite{Reichlova2024} reported a strong anomaly in the temperature dependence of the $c$ lattice parameter in a \MnSi\ film with an abrupt sign change in $dc/dT$ at a temperature that is in between the bulk AFM1 and AFM2 transitions, along with a concomitant anomaly in the resistivity and magnetoresistance.
These features suggest that a magnetic transition with approximately the bulk magnetic energy scale remains present in the films.
Our analysis of the paramagnetic exchange kernel shows that epitaxial constraints imposed on the bulklike stoichiometric \MnSi\ structure do not create 
\emph{another}, more than twice larger, exchange energy scale that could account for the high-temperature anomalous transport.

\section{Conclusions}
\label{sec:conclusions}

Both the AFM2 phase of bulk \MnSi\ and the hypothetical altermagnetic phase of thin \MnSi\ films \cite{Reichlova2024,Rial2024variants} appear naturally in Landau theories. Although both contain the same inversion-even $(1,-1,0)$ Mn2 intracell pattern, they arise from qualitatively different ordering instabilities: bulk AFM2 is a symmetry-protected mode at the $M$ star, whereas the corresponding $\Gamma$-point pattern is a special direction within the collinear sector of the $E_{2g}$ order-parameter manifold.

The paramagnetic DLM exchange kernel identifies the $M$-star AFM2 mode as the dominant instability, and classical Monte Carlo simulations based on this kernel yield an orthogonal $3M$ phase with the magnetic ordering scale comparable to that observed in bulk \MnSi. The selection of the single-$M$ AFM2 phase against the competing orthogonal $3M$ phase in bulk \MnSi\ indicates that negative energetic contributions to the quartic $v_M$ coefficient overcome the positive entropic contribution. Magnetostructural coupling contributes with the required sign, but the homogeneous-strain estimate yields only about a quarter of the competing single-site entropic contribution. Coupling via internal relaxations or separate electronic contributions must therefore provide the remaining stabilization. 

The zone-center $E_{2g}$ mode is substantially weaker compared to the $M$-point instability and, rather than being stabilized, is further suppressed by tensile epitaxial strain that is typical for \MnSi/Si(111) films exhibiting anomalous transport properties \cite{Reichlova2024,Rial2024variants,Badura2025}. 
Nor does such strain generate another exchange instability with a higher energy scale. Thus, the high-temperature Hall-active magnetic state in \MnSi\ films cannot be understood solely as a strain-induced perturbation of stoichiometric bulk \MnSi. This result should motivate further studies of the structure and chemistry specific to the thin-film material exhibiting the elevated N\'eel temperature along with unusual magnetic and transport properties.

\begin{acknowledgments}
I am grateful to Igor Mazin and Daniel Agterberg for useful discussions, and to Vincent Baltz for a careful reading of the manuscript and helpful comments.
This work was supported by the U.S. Department of Energy (DOE) Established Program to Stimulate Competitive Research (EPSCoR) through Grant No. DE-SC0024284. It was completed utilizing the Holland Computing Center of the University of Nebraska, which receives support from the UNL Office of Research and Innovation, and the Nebraska Research Initiative. I used OpenAI GPT-5.6 Sol and GPT-6 Astra models interactively to explore and refine theoretical arguments and derivations, to organize and edit the manuscript, and to generate plotting and data-conversion scripts. I have critically evaluated and verified all model output used in this work.
\end{acknowledgments}


\begin{thebibliography}{47}%
\makeatletter
\providecommand \@ifxundefined [1]{%
 \@ifx{#1\undefined}
}%
\providecommand \@ifnum [1]{%
 \ifnum #1\expandafter \@firstoftwo
 \else \expandafter \@secondoftwo
 \fi
}%
\providecommand \@ifx [1]{%
 \ifx #1\expandafter \@firstoftwo
 \else \expandafter \@secondoftwo
 \fi
}%
\providecommand \natexlab [1]{#1}%
\providecommand \enquote  [1]{``#1''}%
\providecommand \bibnamefont  [1]{#1}%
\providecommand \bibfnamefont [1]{#1}%
\providecommand \citenamefont [1]{#1}%
\providecommand \href@noop [0]{\@secondoftwo}%
\providecommand \href [0]{\begingroup \@sanitize@url \@href}%
\providecommand \@href[1]{\@@startlink{#1}\@@href}%
\providecommand \@@href[1]{\endgroup#1\@@endlink}%
\providecommand \@sanitize@url [0]{\catcode `\\12\catcode `\$12\catcode `\&12\catcode `\#12\catcode `\^12\catcode `\_12\catcode `\%12\relax}%
\providecommand \@@startlink[1]{}%
\providecommand \@@endlink[0]{}%
\providecommand \url  [0]{\begingroup\@sanitize@url \@url }%
\providecommand \@url [1]{\endgroup\@href {#1}{\urlprefix }}%
\providecommand \urlprefix  [0]{URL }%
\providecommand \Eprint [0]{\href }%
\providecommand \doibase [0]{https://doi.org/}%
\providecommand \selectlanguage [0]{\@gobble}%
\providecommand \bibinfo  [0]{\@secondoftwo}%
\providecommand \bibfield  [0]{\@secondoftwo}%
\providecommand \translation [1]{[#1]}%
\providecommand \BibitemOpen [0]{}%
\providecommand \bibitemStop [0]{}%
\providecommand \bibitemNoStop [0]{.\EOS\space}%
\providecommand \EOS [0]{\spacefactor3000\relax}%
\providecommand \BibitemShut  [1]{\csname bibitem#1\endcsname}%
\let\auto@bib@innerbib\@empty
\bibitem [{\citenamefont {Brown}\ \emph {et~al.}(1992)\citenamefont {Brown}, \citenamefont {Forsyth}, \citenamefont {Nunez},\ and\ \citenamefont {Tasset}}]{BrownForsythNunezTasset1992}%
  \BibitemOpen
  \bibfield  {author} {\bibinfo {author} {\bibfnamefont {P.~J.}\ \bibnamefont {Brown}}, \bibinfo {author} {\bibfnamefont {J.~B.}\ \bibnamefont {Forsyth}}, \bibinfo {author} {\bibfnamefont {V.}~\bibnamefont {Nunez}},\ and\ \bibinfo {author} {\bibfnamefont {F.}~\bibnamefont {Tasset}},\ }\bibfield  {title} {\bibinfo {title} {The low-temperature antiferromagnetic structure of {Mn$_5$Si$_3$} revised in the light of neutron polarimetry},\ }\href {https://doi.org/10.1088/0953-8984/4/49/029} {\bibfield  {journal} {\bibinfo  {journal} {J. Phys.: Condens. Matter}\ }\textbf {\bibinfo {volume} {4}},\ \bibinfo {pages} {10025} (\bibinfo {year} {1992})}\BibitemShut {NoStop}%
\bibitem [{\citenamefont {Brown}\ and\ \citenamefont {Forsyth}(1995)}]{BrownForsyth1995}%
  \BibitemOpen
  \bibfield  {author} {\bibinfo {author} {\bibfnamefont {P.~J.}\ \bibnamefont {Brown}}\ and\ \bibinfo {author} {\bibfnamefont {J.~B.}\ \bibnamefont {Forsyth}},\ }\bibfield  {title} {\bibinfo {title} {Antiferromagnetism in {Mn$_5$Si$_3$}: The magnetic structure of the {AF2} phase at 70 {K}},\ }\href {https://doi.org/10.1088/0953-8984/7/39/004} {\bibfield  {journal} {\bibinfo  {journal} {J. Phys.: Condens. Matter}\ }\textbf {\bibinfo {volume} {7}},\ \bibinfo {pages} {7619} (\bibinfo {year} {1995})}\BibitemShut {NoStop}%
\bibitem [{\citenamefont {Reichlov{\'a}}\ \emph {et~al.}(2024)\citenamefont {Reichlov{\'a}}, \citenamefont {Lopes~Seeger}, \citenamefont {Gonz{\'a}lez-Hern{\'a}ndez}, \citenamefont {Kounta}, \citenamefont {Schlitz}, \citenamefont {Kriegner}, \citenamefont {Ritzinger}, \citenamefont {Lammel}, \citenamefont {Leivisk{\"a}}, \citenamefont {Birk~Hellenes}, \citenamefont {Olejn{\'i}k}, \citenamefont {Pet{\v r}{\'i}{\v c}ek}, \citenamefont {Dole{\v z}al}, \citenamefont {Hor{\'a}k}, \citenamefont {Schmoranzerov{\'a}}, \citenamefont {Badura}, \citenamefont {Bertaina}, \citenamefont {Thomas}, \citenamefont {Baltz}, \citenamefont {Michez}, \citenamefont {Sinova}, \citenamefont {Goennenwein}, \citenamefont {Jungwirth},\ and\ \citenamefont {{\v S}mejkal}}]{Reichlova2024}%
  \BibitemOpen
  \bibfield  {author} {\bibinfo {author} {\bibfnamefont {H.}~\bibnamefont {Reichlov{\'a}}}, \bibinfo {author} {\bibfnamefont {R.}~\bibnamefont {Lopes~Seeger}}, \bibinfo {author} {\bibfnamefont {R.}~\bibnamefont {Gonz{\'a}lez-Hern{\'a}ndez}}, \bibinfo {author} {\bibfnamefont {I.}~\bibnamefont {Kounta}}, \bibinfo {author} {\bibfnamefont {R.}~\bibnamefont {Schlitz}}, \bibinfo {author} {\bibfnamefont {D.}~\bibnamefont {Kriegner}}, \bibinfo {author} {\bibfnamefont {P.}~\bibnamefont {Ritzinger}}, \bibinfo {author} {\bibfnamefont {M.}~\bibnamefont {Lammel}}, \bibinfo {author} {\bibfnamefont {M.}~\bibnamefont {Leivisk{\"a}}}, \bibinfo {author} {\bibfnamefont {A.}~\bibnamefont {Birk~Hellenes}}, \bibinfo {author} {\bibfnamefont {K.}~\bibnamefont {Olejn{\'i}k}}, \bibinfo {author} {\bibfnamefont {V.}~\bibnamefont {Pet{\v r}{\'i}{\v c}ek}}, \bibinfo {author} {\bibfnamefont {P.}~\bibnamefont {Dole{\v z}al}}, \bibinfo {author} {\bibfnamefont {L.}~\bibnamefont {Hor{\'a}k}}, \bibinfo {author} {\bibfnamefont {E.}~\bibnamefont
  {Schmoranzerov{\'a}}}, \bibinfo {author} {\bibfnamefont {A.}~\bibnamefont {Badura}}, \bibinfo {author} {\bibfnamefont {S.}~\bibnamefont {Bertaina}}, \bibinfo {author} {\bibfnamefont {A.}~\bibnamefont {Thomas}}, \bibinfo {author} {\bibfnamefont {V.}~\bibnamefont {Baltz}}, \bibinfo {author} {\bibfnamefont {L.}~\bibnamefont {Michez}}, \bibinfo {author} {\bibfnamefont {J.}~\bibnamefont {Sinova}}, \bibinfo {author} {\bibfnamefont {S.~T.~B.}\ \bibnamefont {Goennenwein}}, \bibinfo {author} {\bibfnamefont {T.}~\bibnamefont {Jungwirth}},\ and\ \bibinfo {author} {\bibfnamefont {L.}~\bibnamefont {{\v S}mejkal}},\ }\bibfield  {title} {\bibinfo {title} {Observation of a spontaneous anomalous {Hall} response in the {Mn$_5$Si$_3$} $d$-wave altermagnet candidate},\ }\href {https://doi.org/10.1038/s41467-024-48493-w} {\bibfield  {journal} {\bibinfo  {journal} {Nat. Commun.}\ }\textbf {\bibinfo {volume} {15}},\ \bibinfo {pages} {4961} (\bibinfo {year} {2024})}\BibitemShut {NoStop}%
\bibitem [{\citenamefont {Kounta}\ \emph {et~al.}(2023)\citenamefont {Kounta}, \citenamefont {Reichlova}, \citenamefont {Kriegner}, \citenamefont {Lopes~Seeger}, \citenamefont {Bad'ura}, \citenamefont {Leiviska}, \citenamefont {Boussadi}, \citenamefont {Heresanu}, \citenamefont {Bertaina}, \citenamefont {Petit}, \citenamefont {Schmoranzerova}, \citenamefont {Smejkal}, \citenamefont {Sinova}, \citenamefont {Jungwirth}, \citenamefont {Baltz}, \citenamefont {Goennenwein},\ and\ \citenamefont {Michez}}]{Kounta2023}%
  \BibitemOpen
  \bibfield  {author} {\bibinfo {author} {\bibfnamefont {I.}~\bibnamefont {Kounta}}, \bibinfo {author} {\bibfnamefont {H.}~\bibnamefont {Reichlova}}, \bibinfo {author} {\bibfnamefont {D.}~\bibnamefont {Kriegner}}, \bibinfo {author} {\bibfnamefont {R.}~\bibnamefont {Lopes~Seeger}}, \bibinfo {author} {\bibfnamefont {A.}~\bibnamefont {Bad'ura}}, \bibinfo {author} {\bibfnamefont {M.}~\bibnamefont {Leiviska}}, \bibinfo {author} {\bibfnamefont {A.}~\bibnamefont {Boussadi}}, \bibinfo {author} {\bibfnamefont {V.}~\bibnamefont {Heresanu}}, \bibinfo {author} {\bibfnamefont {S.}~\bibnamefont {Bertaina}}, \bibinfo {author} {\bibfnamefont {M.}~\bibnamefont {Petit}}, \bibinfo {author} {\bibfnamefont {E.}~\bibnamefont {Schmoranzerova}}, \bibinfo {author} {\bibfnamefont {L.}~\bibnamefont {Smejkal}}, \bibinfo {author} {\bibfnamefont {J.}~\bibnamefont {Sinova}}, \bibinfo {author} {\bibfnamefont {T.}~\bibnamefont {Jungwirth}}, \bibinfo {author} {\bibfnamefont {V.}~\bibnamefont {Baltz}}, \bibinfo {author} {\bibfnamefont
  {S.~T.~B.}\ \bibnamefont {Goennenwein}},\ and\ \bibinfo {author} {\bibfnamefont {L.}~\bibnamefont {Michez}},\ }\bibfield  {title} {\bibinfo {title} {Competitive actions of {MnSi} in the epitaxial growth of {${\mathrm{Mn}}_{5}{\mathrm{Si}}_{3}$} thin films on {Si(111)}},\ }\href {https://doi.org/10.1103/PhysRevMaterials.7.024416} {\bibfield  {journal} {\bibinfo  {journal} {Phys. Rev. Mater.}\ }\textbf {\bibinfo {volume} {7}},\ \bibinfo {pages} {024416} (\bibinfo {year} {2023})}\BibitemShut {NoStop}%
\bibitem [{\citenamefont {Leivisk\"a}\ \emph {et~al.}(2024)\citenamefont {Leivisk\"a}, \citenamefont {Rial}, \citenamefont {Bad'ura}, \citenamefont {Seeger}, \citenamefont {Kounta}, \citenamefont {Beckert}, \citenamefont {Kriegner}, \citenamefont {Joumard}, \citenamefont {Schmoranzerov\'a}, \citenamefont {Sinova}, \citenamefont {Gomonay}, \citenamefont {Thomas}, \citenamefont {Goennenwein}, \citenamefont {Reichlov\'a}, \citenamefont {\ifmmode~\check{S}\else \v{S}\fi{}mejkal}, \citenamefont {Michez}, \citenamefont {Jungwirth},\ and\ \citenamefont {Baltz}}]{Leiviska2024}%
  \BibitemOpen
  \bibfield  {author} {\bibinfo {author} {\bibfnamefont {M.}~\bibnamefont {Leivisk\"a}}, \bibinfo {author} {\bibfnamefont {J.}~\bibnamefont {Rial}}, \bibinfo {author} {\bibfnamefont {A.}~\bibnamefont {Bad'ura}}, \bibinfo {author} {\bibfnamefont {R.~L.}\ \bibnamefont {Seeger}}, \bibinfo {author} {\bibfnamefont {I.}~\bibnamefont {Kounta}}, \bibinfo {author} {\bibfnamefont {S.}~\bibnamefont {Beckert}}, \bibinfo {author} {\bibfnamefont {D.}~\bibnamefont {Kriegner}}, \bibinfo {author} {\bibfnamefont {I.}~\bibnamefont {Joumard}}, \bibinfo {author} {\bibfnamefont {E.}~\bibnamefont {Schmoranzerov\'a}}, \bibinfo {author} {\bibfnamefont {J.}~\bibnamefont {Sinova}}, \bibinfo {author} {\bibfnamefont {O.}~\bibnamefont {Gomonay}}, \bibinfo {author} {\bibfnamefont {A.}~\bibnamefont {Thomas}}, \bibinfo {author} {\bibfnamefont {S.~T.~B.}\ \bibnamefont {Goennenwein}}, \bibinfo {author} {\bibfnamefont {H.}~\bibnamefont {Reichlov\'a}}, \bibinfo {author} {\bibfnamefont {L.}~\bibnamefont {\ifmmode~\check{S}\else
  \v{S}\fi{}mejkal}}, \bibinfo {author} {\bibfnamefont {L.}~\bibnamefont {Michez}}, \bibinfo {author} {\bibfnamefont {T.}~\bibnamefont {Jungwirth}},\ and\ \bibinfo {author} {\bibfnamefont {V.}~\bibnamefont {Baltz}},\ }\bibfield  {title} {\bibinfo {title} {Anisotropy of the anomalous {Hall} effect in thin films of the altermagnet candidate {${\mathrm{Mn}}_{5}{\mathrm{Si}}_{3}$}},\ }\href {https://doi.org/10.1103/PhysRevB.109.224430} {\bibfield  {journal} {\bibinfo  {journal} {Phys. Rev. B}\ }\textbf {\bibinfo {volume} {109}},\ \bibinfo {pages} {224430} (\bibinfo {year} {2024})}\BibitemShut {NoStop}%
\bibitem [{\citenamefont {Han}\ \emph {et~al.}(2024)\citenamefont {Han}, \citenamefont {Fu}, \citenamefont {Peng}, \citenamefont {Cheng}, \citenamefont {Dai}, \citenamefont {Liu}, \citenamefont {Li}, \citenamefont {Zhang}, \citenamefont {Zhu}, \citenamefont {Bai}, \citenamefont {Zhou}, \citenamefont {Liang}, \citenamefont {Chen}, \citenamefont {Wang}, \citenamefont {Chen}, \citenamefont {Yang}, \citenamefont {Zhang}, \citenamefont {Song}, \citenamefont {Liu},\ and\ \citenamefont {Pan}}]{Han2024SciAdv}%
  \BibitemOpen
  \bibfield  {author} {\bibinfo {author} {\bibfnamefont {L.}~\bibnamefont {Han}}, \bibinfo {author} {\bibfnamefont {X.}~\bibnamefont {Fu}}, \bibinfo {author} {\bibfnamefont {R.}~\bibnamefont {Peng}}, \bibinfo {author} {\bibfnamefont {X.}~\bibnamefont {Cheng}}, \bibinfo {author} {\bibfnamefont {J.}~\bibnamefont {Dai}}, \bibinfo {author} {\bibfnamefont {L.}~\bibnamefont {Liu}}, \bibinfo {author} {\bibfnamefont {Y.}~\bibnamefont {Li}}, \bibinfo {author} {\bibfnamefont {Y.}~\bibnamefont {Zhang}}, \bibinfo {author} {\bibfnamefont {W.}~\bibnamefont {Zhu}}, \bibinfo {author} {\bibfnamefont {H.}~\bibnamefont {Bai}}, \bibinfo {author} {\bibfnamefont {Y.}~\bibnamefont {Zhou}}, \bibinfo {author} {\bibfnamefont {S.}~\bibnamefont {Liang}}, \bibinfo {author} {\bibfnamefont {C.}~\bibnamefont {Chen}}, \bibinfo {author} {\bibfnamefont {Q.}~\bibnamefont {Wang}}, \bibinfo {author} {\bibfnamefont {X.}~\bibnamefont {Chen}}, \bibinfo {author} {\bibfnamefont {L.}~\bibnamefont {Yang}}, \bibinfo {author} {\bibfnamefont
  {Y.}~\bibnamefont {Zhang}}, \bibinfo {author} {\bibfnamefont {C.}~\bibnamefont {Song}}, \bibinfo {author} {\bibfnamefont {J.}~\bibnamefont {Liu}},\ and\ \bibinfo {author} {\bibfnamefont {F.}~\bibnamefont {Pan}},\ }\bibfield  {title} {\bibinfo {title} {Electrical $180^\circ$ switching of {N\'eel} vector in spin-splitting antiferromagnet},\ }\href {https://doi.org/10.1126/sciadv.adn0479} {\bibfield  {journal} {\bibinfo  {journal} {Sci. Adv.}\ }\textbf {\bibinfo {volume} {10}},\ \bibinfo {pages} {eadn0479} (\bibinfo {year} {2024})}\BibitemShut {NoStop}%
\bibitem [{\citenamefont {Badura}\ \emph {et~al.}(2025)\citenamefont {Badura}, \citenamefont {Campos}, \citenamefont {Bharadwaj}, \citenamefont {Kounta}, \citenamefont {Michez}, \citenamefont {Petit}, \citenamefont {Rial}, \citenamefont {Leiviska}, \citenamefont {Baltz}, \citenamefont {Krizek}, \citenamefont {Kriegner}, \citenamefont {Zelezny}, \citenamefont {Zemen}, \citenamefont {Telkamp}, \citenamefont {Sailler}, \citenamefont {Lammel}, \citenamefont {Jaeschke-Ubiergo}, \citenamefont {Hellenes}, \citenamefont {Gonzalez-Hernandez}, \citenamefont {Sinova}, \citenamefont {Jungwirth}, \citenamefont {Goennenwein}, \citenamefont {Smejkal},\ and\ \citenamefont {Reichlova}}]{Badura2025}%
  \BibitemOpen
  \bibfield  {author} {\bibinfo {author} {\bibfnamefont {A.}~\bibnamefont {Badura}}, \bibinfo {author} {\bibfnamefont {W.~H.}\ \bibnamefont {Campos}}, \bibinfo {author} {\bibfnamefont {V.~K.}\ \bibnamefont {Bharadwaj}}, \bibinfo {author} {\bibfnamefont {I.}~\bibnamefont {Kounta}}, \bibinfo {author} {\bibfnamefont {L.}~\bibnamefont {Michez}}, \bibinfo {author} {\bibfnamefont {M.}~\bibnamefont {Petit}}, \bibinfo {author} {\bibfnamefont {J.}~\bibnamefont {Rial}}, \bibinfo {author} {\bibfnamefont {M.}~\bibnamefont {Leiviska}}, \bibinfo {author} {\bibfnamefont {V.}~\bibnamefont {Baltz}}, \bibinfo {author} {\bibfnamefont {F.}~\bibnamefont {Krizek}}, \bibinfo {author} {\bibfnamefont {D.}~\bibnamefont {Kriegner}}, \bibinfo {author} {\bibfnamefont {J.}~\bibnamefont {Zelezny}}, \bibinfo {author} {\bibfnamefont {J.}~\bibnamefont {Zemen}}, \bibinfo {author} {\bibfnamefont {S.}~\bibnamefont {Telkamp}}, \bibinfo {author} {\bibfnamefont {S.}~\bibnamefont {Sailler}}, \bibinfo {author} {\bibfnamefont {M.}~\bibnamefont
  {Lammel}}, \bibinfo {author} {\bibfnamefont {R.}~\bibnamefont {Jaeschke-Ubiergo}}, \bibinfo {author} {\bibfnamefont {A.~B.}\ \bibnamefont {Hellenes}}, \bibinfo {author} {\bibfnamefont {R.}~\bibnamefont {Gonzalez-Hernandez}}, \bibinfo {author} {\bibfnamefont {J.}~\bibnamefont {Sinova}}, \bibinfo {author} {\bibfnamefont {T.}~\bibnamefont {Jungwirth}}, \bibinfo {author} {\bibfnamefont {S.~T.~B.}\ \bibnamefont {Goennenwein}}, \bibinfo {author} {\bibfnamefont {L.}~\bibnamefont {Smejkal}},\ and\ \bibinfo {author} {\bibfnamefont {H.}~\bibnamefont {Reichlova}},\ }\bibfield  {title} {\bibinfo {title} {Observation of the anomalous {Nernst} effect in altermagnetic candidate {Mn$_5$Si$_3$}},\ }\href {https://doi.org/10.1038/s41467-025-62331-7} {\bibfield  {journal} {\bibinfo  {journal} {Nat. Commun.}\ }\textbf {\bibinfo {volume} {16}},\ \bibinfo {pages} {7111} (\bibinfo {year} {2025})}\BibitemShut {NoStop}%
\bibitem [{\citenamefont {Han}\ \emph {et~al.}(2025{\natexlab{a}})\citenamefont {Han}, \citenamefont {Fu}, \citenamefont {He}, \citenamefont {Dai}, \citenamefont {Zhu}, \citenamefont {Yang}, \citenamefont {Chen}, \citenamefont {Zhang}, \citenamefont {Zhu}, \citenamefont {Bai}, \citenamefont {Chen}, \citenamefont {Hou}, \citenamefont {Wan}, \citenamefont {Han}, \citenamefont {Song}, \citenamefont {Liu},\ and\ \citenamefont {Pan}}]{Han2025PRA}%
  \BibitemOpen
  \bibfield  {author} {\bibinfo {author} {\bibfnamefont {L.}~\bibnamefont {Han}}, \bibinfo {author} {\bibfnamefont {X.}~\bibnamefont {Fu}}, \bibinfo {author} {\bibfnamefont {W.}~\bibnamefont {He}}, \bibinfo {author} {\bibfnamefont {J.}~\bibnamefont {Dai}}, \bibinfo {author} {\bibfnamefont {Y.}~\bibnamefont {Zhu}}, \bibinfo {author} {\bibfnamefont {W.}~\bibnamefont {Yang}}, \bibinfo {author} {\bibfnamefont {Y.}~\bibnamefont {Chen}}, \bibinfo {author} {\bibfnamefont {J.}~\bibnamefont {Zhang}}, \bibinfo {author} {\bibfnamefont {W.}~\bibnamefont {Zhu}}, \bibinfo {author} {\bibfnamefont {H.}~\bibnamefont {Bai}}, \bibinfo {author} {\bibfnamefont {C.}~\bibnamefont {Chen}}, \bibinfo {author} {\bibfnamefont {D.}~\bibnamefont {Hou}}, \bibinfo {author} {\bibfnamefont {C.}~\bibnamefont {Wan}}, \bibinfo {author} {\bibfnamefont {X.}~\bibnamefont {Han}}, \bibinfo {author} {\bibfnamefont {C.}~\bibnamefont {Song}}, \bibinfo {author} {\bibfnamefont {J.}~\bibnamefont {Liu}},\ and\ \bibinfo {author} {\bibfnamefont
  {F.}~\bibnamefont {Pan}},\ }\bibfield  {title} {\bibinfo {title} {Nonvolatile anomalous {Nernst} effect in {${\mathrm{Mn}}_{5}{\mathrm{Si}}_{3}$} with a collinear {N\'eel} vector},\ }\href {https://doi.org/10.1103/PhysRevApplied.23.044066} {\bibfield  {journal} {\bibinfo  {journal} {Phys. Rev. Appl.}\ }\textbf {\bibinfo {volume} {23}},\ \bibinfo {pages} {044066} (\bibinfo {year} {2025}{\natexlab{a}})}\BibitemShut {NoStop}%
\bibitem [{\citenamefont {Rial}\ \emph {et~al.}(2024)\citenamefont {Rial}, \citenamefont {Leivisk\"a}, \citenamefont {Skobjin}, \citenamefont {Bad'ura}, \citenamefont {Gaudin}, \citenamefont {Disdier}, \citenamefont {Schlitz}, \citenamefont {Kounta}, \citenamefont {Beckert}, \citenamefont {Kriegner}, \citenamefont {Thomas}, \citenamefont {Schmoranzerov\'a}, \citenamefont {\ifmmode~\check{S}\else \v{S}\fi{}mejkal}, \citenamefont {Sinova}, \citenamefont {Jungwirth}, \citenamefont {Michez}, \citenamefont {Reichlov\'a}, \citenamefont {Goennenwein}, \citenamefont {Gomonay},\ and\ \citenamefont {Baltz}}]{Rial2024variants}%
  \BibitemOpen
  \bibfield  {author} {\bibinfo {author} {\bibfnamefont {J.}~\bibnamefont {Rial}}, \bibinfo {author} {\bibfnamefont {M.}~\bibnamefont {Leivisk\"a}}, \bibinfo {author} {\bibfnamefont {G.}~\bibnamefont {Skobjin}}, \bibinfo {author} {\bibfnamefont {A.}~\bibnamefont {Bad'ura}}, \bibinfo {author} {\bibfnamefont {G.}~\bibnamefont {Gaudin}}, \bibinfo {author} {\bibfnamefont {F.}~\bibnamefont {Disdier}}, \bibinfo {author} {\bibfnamefont {R.}~\bibnamefont {Schlitz}}, \bibinfo {author} {\bibfnamefont {I.}~\bibnamefont {Kounta}}, \bibinfo {author} {\bibfnamefont {S.}~\bibnamefont {Beckert}}, \bibinfo {author} {\bibfnamefont {D.}~\bibnamefont {Kriegner}}, \bibinfo {author} {\bibfnamefont {A.}~\bibnamefont {Thomas}}, \bibinfo {author} {\bibfnamefont {E.}~\bibnamefont {Schmoranzerov\'a}}, \bibinfo {author} {\bibfnamefont {L.}~\bibnamefont {\ifmmode~\check{S}\else \v{S}\fi{}mejkal}}, \bibinfo {author} {\bibfnamefont {J.}~\bibnamefont {Sinova}}, \bibinfo {author} {\bibfnamefont {T.}~\bibnamefont {Jungwirth}}, \bibinfo
  {author} {\bibfnamefont {L.}~\bibnamefont {Michez}}, \bibinfo {author} {\bibfnamefont {H.}~\bibnamefont {Reichlov\'a}}, \bibinfo {author} {\bibfnamefont {S.~T.~B.}\ \bibnamefont {Goennenwein}}, \bibinfo {author} {\bibfnamefont {O.}~\bibnamefont {Gomonay}},\ and\ \bibinfo {author} {\bibfnamefont {V.}~\bibnamefont {Baltz}},\ }\bibfield  {title} {\bibinfo {title} {Altermagnetic variants in thin films of $\mathrm{M}{\mathrm{n}}_{5}\mathrm{S}{\mathrm{i}}_{3}$},\ }\href {https://doi.org/10.1103/PhysRevB.110.L220411} {\bibfield  {journal} {\bibinfo  {journal} {Phys. Rev. B}\ }\textbf {\bibinfo {volume} {110}},\ \bibinfo {pages} {L220411} (\bibinfo {year} {2024})}\BibitemShut {NoStop}%
\bibitem [{\citenamefont {Han}\ \emph {et~al.}(2025{\natexlab{b}})\citenamefont {Han}, \citenamefont {Fu}, \citenamefont {Song}, \citenamefont {Zhu}, \citenamefont {Li}, \citenamefont {Zhu}, \citenamefont {Bai}, \citenamefont {Chu}, \citenamefont {Dai}, \citenamefont {Liang}, \citenamefont {Sawicki}, \citenamefont {Liu},\ and\ \citenamefont {Pan}}]{han2025}%
  \BibitemOpen
  \bibfield  {author} {\bibinfo {author} {\bibfnamefont {L.}~\bibnamefont {Han}}, \bibinfo {author} {\bibfnamefont {X.}~\bibnamefont {Fu}}, \bibinfo {author} {\bibfnamefont {C.}~\bibnamefont {Song}}, \bibinfo {author} {\bibfnamefont {Y.}~\bibnamefont {Zhu}}, \bibinfo {author} {\bibfnamefont {X.}~\bibnamefont {Li}}, \bibinfo {author} {\bibfnamefont {Z.}~\bibnamefont {Zhu}}, \bibinfo {author} {\bibfnamefont {H.}~\bibnamefont {Bai}}, \bibinfo {author} {\bibfnamefont {R.}~\bibnamefont {Chu}}, \bibinfo {author} {\bibfnamefont {J.}~\bibnamefont {Dai}}, \bibinfo {author} {\bibfnamefont {S.}~\bibnamefont {Liang}}, \bibinfo {author} {\bibfnamefont {M.}~\bibnamefont {Sawicki}}, \bibinfo {author} {\bibfnamefont {J.}~\bibnamefont {Liu}},\ and\ \bibinfo {author} {\bibfnamefont {F.}~\bibnamefont {Pan}},\ }\href {https://arxiv.org/abs/2502.04920} {\bibinfo {title} {Discovery of a large magnetic nonlinear {Hall} effect in an altermagnet}} (\bibinfo {year} {2025}{\natexlab{b}}),\ \Eprint {https://arxiv.org/abs/2502.04920}
  {arXiv:2502.04920} \BibitemShut {NoStop}%
\bibitem [{\citenamefont {Skobjin}\ \emph {et~al.}(2026)\citenamefont {Skobjin}, \citenamefont {Rial}, \citenamefont {Beckert}, \citenamefont {Reichlova}, \citenamefont {Baltz}, \citenamefont {Michez}, \citenamefont {Schlitz}, \citenamefont {Lammel},\ and\ \citenamefont {Goennenwein}}]{Skobjin2026}%
  \BibitemOpen
  \bibfield  {author} {\bibinfo {author} {\bibfnamefont {G.}~\bibnamefont {Skobjin}}, \bibinfo {author} {\bibfnamefont {J.}~\bibnamefont {Rial}}, \bibinfo {author} {\bibfnamefont {S.}~\bibnamefont {Beckert}}, \bibinfo {author} {\bibfnamefont {H.}~\bibnamefont {Reichlova}}, \bibinfo {author} {\bibfnamefont {V.}~\bibnamefont {Baltz}}, \bibinfo {author} {\bibfnamefont {L.}~\bibnamefont {Michez}}, \bibinfo {author} {\bibfnamefont {R.}~\bibnamefont {Schlitz}}, \bibinfo {author} {\bibfnamefont {M.}~\bibnamefont {Lammel}},\ and\ \bibinfo {author} {\bibfnamefont {S.~T.~B.}\ \bibnamefont {Goennenwein}},\ }\bibfield  {title} {\bibinfo {title} {Magnetic aftereffect and {Barkhausen} jumps in thin altermagnetic {Mn$_5$Si$_3$} films},\ }\href {https://doi.org/10.1063/5.0314005} {\bibfield  {journal} {\bibinfo  {journal} {Appl. Phys. Lett.}\ }\textbf {\bibinfo {volume} {128}},\ \bibinfo {pages} {102407} (\bibinfo {year} {2026})}\BibitemShut {NoStop}%
\bibitem [{\citenamefont {Mencos}\ \emph {et~al.}(2025)\citenamefont {Mencos}, \citenamefont {Badura}, \citenamefont {Dolan}, \citenamefont {Beckert}, \citenamefont {Gonzalez-Hernandez}, \citenamefont {Kounta}, \citenamefont {Petit}, \citenamefont {Guillemard}, \citenamefont {Hellenes}, \citenamefont {Campos}, \citenamefont {Rial}, \citenamefont {Kriegner}, \citenamefont {Baltz}, \citenamefont {Hueso}, \citenamefont {Sinova}, \citenamefont {Gomonay}, \citenamefont {Jungwirth}, \citenamefont {Smejkal}, \citenamefont {Michez}, \citenamefont {Reichlova},\ and\ \citenamefont {Casanova}}]{mencos2025}%
  \BibitemOpen
  \bibfield  {author} {\bibinfo {author} {\bibfnamefont {J.}~\bibnamefont {Mencos}}, \bibinfo {author} {\bibfnamefont {A.}~\bibnamefont {Badura}}, \bibinfo {author} {\bibfnamefont {E.}~\bibnamefont {Dolan}}, \bibinfo {author} {\bibfnamefont {S.}~\bibnamefont {Beckert}}, \bibinfo {author} {\bibfnamefont {R.}~\bibnamefont {Gonzalez-Hernandez}}, \bibinfo {author} {\bibfnamefont {I.}~\bibnamefont {Kounta}}, \bibinfo {author} {\bibfnamefont {M.}~\bibnamefont {Petit}}, \bibinfo {author} {\bibfnamefont {C.}~\bibnamefont {Guillemard}}, \bibinfo {author} {\bibfnamefont {A.~B.}\ \bibnamefont {Hellenes}}, \bibinfo {author} {\bibfnamefont {W.}~\bibnamefont {Campos}}, \bibinfo {author} {\bibfnamefont {J.}~\bibnamefont {Rial}}, \bibinfo {author} {\bibfnamefont {D.}~\bibnamefont {Kriegner}}, \bibinfo {author} {\bibfnamefont {V.}~\bibnamefont {Baltz}}, \bibinfo {author} {\bibfnamefont {L.~E.}\ \bibnamefont {Hueso}}, \bibinfo {author} {\bibfnamefont {J.}~\bibnamefont {Sinova}}, \bibinfo {author} {\bibfnamefont
  {O.}~\bibnamefont {Gomonay}}, \bibinfo {author} {\bibfnamefont {T.}~\bibnamefont {Jungwirth}}, \bibinfo {author} {\bibfnamefont {L.}~\bibnamefont {Smejkal}}, \bibinfo {author} {\bibfnamefont {L.}~\bibnamefont {Michez}}, \bibinfo {author} {\bibfnamefont {H.}~\bibnamefont {Reichlova}},\ and\ \bibinfo {author} {\bibfnamefont {F.}~\bibnamefont {Casanova}},\ }\href {https://arxiv.org/abs/2512.17427} {\bibinfo {title} {Direct demonstration of time-reversal-symmetry-breaking spin injection from a compensated magnet}} (\bibinfo {year} {2025}),\ \Eprint {https://arxiv.org/abs/2512.17427} {arXiv:2512.17427} \BibitemShut {NoStop}%
\bibitem [{\citenamefont {\ifmmode~\check{S}\else \v{S}\fi{}mejkal}\ \emph {et~al.}(2022{\natexlab{a}})\citenamefont {\ifmmode~\check{S}\else \v{S}\fi{}mejkal}, \citenamefont {Sinova},\ and\ \citenamefont {Jungwirth}}]{Smejkal1}%
  \BibitemOpen
  \bibfield  {author} {\bibinfo {author} {\bibfnamefont {L.}~\bibnamefont {\ifmmode~\check{S}\else \v{S}\fi{}mejkal}}, \bibinfo {author} {\bibfnamefont {J.}~\bibnamefont {Sinova}},\ and\ \bibinfo {author} {\bibfnamefont {T.}~\bibnamefont {Jungwirth}},\ }\bibfield  {title} {\bibinfo {title} {Beyond conventional ferromagnetism and antiferromagnetism: A phase with nonrelativistic spin and crystal rotation symmetry},\ }\href {https://doi.org/10.1103/PhysRevX.12.031042} {\bibfield  {journal} {\bibinfo  {journal} {Phys. Rev. X}\ }\textbf {\bibinfo {volume} {12}},\ \bibinfo {pages} {031042} (\bibinfo {year} {2022}{\natexlab{a}})}\BibitemShut {NoStop}%
\bibitem [{\citenamefont {\ifmmode~\check{S}\else \v{S}\fi{}mejkal}\ \emph {et~al.}(2022{\natexlab{b}})\citenamefont {\ifmmode~\check{S}\else \v{S}\fi{}mejkal}, \citenamefont {Sinova},\ and\ \citenamefont {Jungwirth}}]{Smejkal2}%
  \BibitemOpen
  \bibfield  {author} {\bibinfo {author} {\bibfnamefont {L.}~\bibnamefont {\ifmmode~\check{S}\else \v{S}\fi{}mejkal}}, \bibinfo {author} {\bibfnamefont {J.}~\bibnamefont {Sinova}},\ and\ \bibinfo {author} {\bibfnamefont {T.}~\bibnamefont {Jungwirth}},\ }\bibfield  {title} {\bibinfo {title} {Emerging research landscape of altermagnetism},\ }\href {https://doi.org/10.1103/PhysRevX.12.040501} {\bibfield  {journal} {\bibinfo  {journal} {Phys. Rev. X}\ }\textbf {\bibinfo {volume} {12}},\ \bibinfo {pages} {040501} (\bibinfo {year} {2022}{\natexlab{b}})}\BibitemShut {NoStop}%
\bibitem [{\citenamefont {Song}\ \emph {et~al.}(2025)\citenamefont {Song}, \citenamefont {Bai}, \citenamefont {Zhou}, \citenamefont {Han}, \citenamefont {Reichlova}, \citenamefont {Dil}, \citenamefont {Liu}, \citenamefont {Chen},\ and\ \citenamefont {Pan}}]{Song_review_2025}%
  \BibitemOpen
  \bibfield  {author} {\bibinfo {author} {\bibfnamefont {C.}~\bibnamefont {Song}}, \bibinfo {author} {\bibfnamefont {H.}~\bibnamefont {Bai}}, \bibinfo {author} {\bibfnamefont {Z.}~\bibnamefont {Zhou}}, \bibinfo {author} {\bibfnamefont {L.}~\bibnamefont {Han}}, \bibinfo {author} {\bibfnamefont {H.}~\bibnamefont {Reichlova}}, \bibinfo {author} {\bibfnamefont {J.~H.}\ \bibnamefont {Dil}}, \bibinfo {author} {\bibfnamefont {J.}~\bibnamefont {Liu}}, \bibinfo {author} {\bibfnamefont {X.}~\bibnamefont {Chen}},\ and\ \bibinfo {author} {\bibfnamefont {F.}~\bibnamefont {Pan}},\ }\bibfield  {title} {\bibinfo {title} {Altermagnets as a new class of functional materials},\ }\href {https://doi.org/10.1038/s41578-025-00779-1} {\bibfield  {journal} {\bibinfo  {journal} {Nat. Rev. Mater.}\ }\textbf {\bibinfo {volume} {10}},\ \bibinfo {pages} {473} (\bibinfo {year} {2025})}\BibitemShut {NoStop}%
\bibitem [{\citenamefont {Jungwirth}\ \emph {et~al.}(2026)\citenamefont {Jungwirth}, \citenamefont {Sinova}, \citenamefont {Wadley}, \citenamefont {Kriegner}, \citenamefont {Reichlova}, \citenamefont {Krizek}, \citenamefont {Ohno},\ and\ \citenamefont {Smejkal}}]{Jungwirth2026}%
  \BibitemOpen
  \bibfield  {author} {\bibinfo {author} {\bibfnamefont {T.}~\bibnamefont {Jungwirth}}, \bibinfo {author} {\bibfnamefont {J.}~\bibnamefont {Sinova}}, \bibinfo {author} {\bibfnamefont {P.}~\bibnamefont {Wadley}}, \bibinfo {author} {\bibfnamefont {D.}~\bibnamefont {Kriegner}}, \bibinfo {author} {\bibfnamefont {H.}~\bibnamefont {Reichlova}}, \bibinfo {author} {\bibfnamefont {F.}~\bibnamefont {Krizek}}, \bibinfo {author} {\bibfnamefont {H.}~\bibnamefont {Ohno}},\ and\ \bibinfo {author} {\bibfnamefont {L.}~\bibnamefont {Smejkal}},\ }\bibfield  {title} {\bibinfo {title} {Altermagnetic spintronics},\ }\href {https://doi.org/10.1038/s41567-026-03337-w} {\bibfield  {journal} {\bibinfo  {journal} {Nat. Phys.}\ }\textbf {\bibinfo {volume} {22}},\ \bibinfo {pages} {1012} (\bibinfo {year} {2026})}\BibitemShut {NoStop}%
\bibitem [{\citenamefont {dos Santos}\ \emph {et~al.}(2021)\citenamefont {dos Santos}, \citenamefont {Biniskos}, \citenamefont {Raymond}, \citenamefont {Schmalzl}, \citenamefont {dos Santos~Dias}, \citenamefont {Steffens}, \citenamefont {Persson}, \citenamefont {Bl\"ugel}, \citenamefont {Lounis},\ and\ \citenamefont {Br\"uckel}}]{dosSantos2021}%
  \BibitemOpen
  \bibfield  {author} {\bibinfo {author} {\bibfnamefont {F.~J.}\ \bibnamefont {dos Santos}}, \bibinfo {author} {\bibfnamefont {N.}~\bibnamefont {Biniskos}}, \bibinfo {author} {\bibfnamefont {S.}~\bibnamefont {Raymond}}, \bibinfo {author} {\bibfnamefont {K.}~\bibnamefont {Schmalzl}}, \bibinfo {author} {\bibfnamefont {M.}~\bibnamefont {dos Santos~Dias}}, \bibinfo {author} {\bibfnamefont {P.}~\bibnamefont {Steffens}}, \bibinfo {author} {\bibfnamefont {J.}~\bibnamefont {Persson}}, \bibinfo {author} {\bibfnamefont {S.}~\bibnamefont {Bl\"ugel}}, \bibinfo {author} {\bibfnamefont {S.}~\bibnamefont {Lounis}},\ and\ \bibinfo {author} {\bibfnamefont {T.}~\bibnamefont {Br\"uckel}},\ }\bibfield  {title} {\bibinfo {title} {Spin waves in the collinear antiferromagnetic phase of {${\mathrm{Mn}}_{5}{\mathrm{Si}}_{3}$}},\ }\href {https://doi.org/10.1103/PhysRevB.103.024407} {\bibfield  {journal} {\bibinfo  {journal} {Phys. Rev. B}\ }\textbf {\bibinfo {volume} {103}},\ \bibinfo {pages} {024407} (\bibinfo {year}
  {2021})}\BibitemShut {NoStop}%
\bibitem [{\citenamefont {Corbett}\ \emph {et~al.}(1998)\citenamefont {Corbett}, \citenamefont {Garcia}, \citenamefont {Guloy}, \citenamefont {Hurng}, \citenamefont {Kwon},\ and\ \citenamefont {Leon-Escamilla}}]{Corbett1998}%
  \BibitemOpen
  \bibfield  {author} {\bibinfo {author} {\bibfnamefont {J.~D.}\ \bibnamefont {Corbett}}, \bibinfo {author} {\bibfnamefont {E.}~\bibnamefont {Garcia}}, \bibinfo {author} {\bibfnamefont {A.~M.}\ \bibnamefont {Guloy}}, \bibinfo {author} {\bibfnamefont {W.-M.}\ \bibnamefont {Hurng}}, \bibinfo {author} {\bibfnamefont {Y.-U.}\ \bibnamefont {Kwon}},\ and\ \bibinfo {author} {\bibfnamefont {E.~A.}\ \bibnamefont {Leon-Escamilla}},\ }\bibfield  {title} {\bibinfo {title} {Widespread interstitial chemistry of {Mn$_5$Si$_3$}-type and related phases. {Hidden} impurities and opportunities},\ }\href {https://doi.org/10.1021/cm980223c} {\bibfield  {journal} {\bibinfo  {journal} {Chem. Mater.}\ }\textbf {\bibinfo {volume} {10}},\ \bibinfo {pages} {2824} (\bibinfo {year} {1998})}\BibitemShut {NoStop}%
\bibitem [{\citenamefont {Gajdzik}\ \emph {et~al.}(2000)\citenamefont {Gajdzik}, \citenamefont {Sürgers}, \citenamefont {Kelemen},\ and\ \citenamefont {Löhneysen}}]{Gajdzik2000}%
  \BibitemOpen
  \bibfield  {author} {\bibinfo {author} {\bibfnamefont {M.}~\bibnamefont {Gajdzik}}, \bibinfo {author} {\bibfnamefont {C.}~\bibnamefont {Sürgers}}, \bibinfo {author} {\bibfnamefont {M.}~\bibnamefont {Kelemen}},\ and\ \bibinfo {author} {\bibfnamefont {H.~v.}\ \bibnamefont {Löhneysen}},\ }\bibfield  {title} {\bibinfo {title} {Ferromagnetism in carbon-doped {Mn$_5$Si$_3$} films},\ }\href {https://doi.org/10.1063/1.372597} {\bibfield  {journal} {\bibinfo  {journal} {J. Appl. Phys.}\ }\textbf {\bibinfo {volume} {87}},\ \bibinfo {pages} {6013} (\bibinfo {year} {2000})}\BibitemShut {NoStop}%
\bibitem [{\citenamefont {S{\"u}rgers}\ \emph {et~al.}(2014)\citenamefont {S{\"u}rgers}, \citenamefont {Fischer}, \citenamefont {Winkel},\ and\ \citenamefont {v.~L{\"o}hneysen}}]{Surgers2014}%
  \BibitemOpen
  \bibfield  {author} {\bibinfo {author} {\bibfnamefont {C.}~\bibnamefont {S{\"u}rgers}}, \bibinfo {author} {\bibfnamefont {G.}~\bibnamefont {Fischer}}, \bibinfo {author} {\bibfnamefont {P.}~\bibnamefont {Winkel}},\ and\ \bibinfo {author} {\bibfnamefont {H.}~\bibnamefont {v.~L{\"o}hneysen}},\ }\bibfield  {title} {\bibinfo {title} {Large topological {H}all effect in the non-collinear phase of an antiferromagnet},\ }\href {https://doi.org/10.1038/ncomms4400} {\bibfield  {journal} {\bibinfo  {journal} {Nat. Commun.}\ }\textbf {\bibinfo {volume} {5}},\ \bibinfo {pages} {3400} (\bibinfo {year} {2014})}\BibitemShut {NoStop}%
\bibitem [{\citenamefont {Suergers}\ \emph {et~al.}(2024)\citenamefont {Suergers}, \citenamefont {Fischer}, \citenamefont {Campos}, \citenamefont {Hellenes}, \citenamefont {Smejkal}, \citenamefont {Sinova}, \citenamefont {Merz}, \citenamefont {Wolf},\ and\ \citenamefont {Wernsdorfer}}]{Surgers2024}%
  \BibitemOpen
  \bibfield  {author} {\bibinfo {author} {\bibfnamefont {C.}~\bibnamefont {Suergers}}, \bibinfo {author} {\bibfnamefont {G.}~\bibnamefont {Fischer}}, \bibinfo {author} {\bibfnamefont {W.~H.}\ \bibnamefont {Campos}}, \bibinfo {author} {\bibfnamefont {A.~B.}\ \bibnamefont {Hellenes}}, \bibinfo {author} {\bibfnamefont {L.}~\bibnamefont {Smejkal}}, \bibinfo {author} {\bibfnamefont {J.}~\bibnamefont {Sinova}}, \bibinfo {author} {\bibfnamefont {M.}~\bibnamefont {Merz}}, \bibinfo {author} {\bibfnamefont {T.}~\bibnamefont {Wolf}},\ and\ \bibinfo {author} {\bibfnamefont {W.}~\bibnamefont {Wernsdorfer}},\ }\bibfield  {title} {\bibinfo {title} {Anomalous nernst effect in the noncollinear antiferromagnet mn5si3},\ }\bibfield  {journal} {\bibinfo  {journal} {Commun. Mater.}\ }\textbf {\bibinfo {volume} {5}},\ \href {https://doi.org/10.1038/s43246-024-00617-x} {10.1038/s43246-024-00617-x} (\bibinfo {year} {2024})\BibitemShut {NoStop}%
\bibitem [{\citenamefont {Biniskos}\ \emph {et~al.}(2022)\citenamefont {Biniskos}, \citenamefont {dos Santos}, \citenamefont {Schmalzl}, \citenamefont {Raymond}, \citenamefont {dos Santos~Dias}, \citenamefont {Persson}, \citenamefont {Marzari}, \citenamefont {Bl\"ugel}, \citenamefont {Lounis},\ and\ \citenamefont {Br\"uckel}}]{Biniskos2022}%
  \BibitemOpen
  \bibfield  {author} {\bibinfo {author} {\bibfnamefont {N.}~\bibnamefont {Biniskos}}, \bibinfo {author} {\bibfnamefont {F.~J.}\ \bibnamefont {dos Santos}}, \bibinfo {author} {\bibfnamefont {K.}~\bibnamefont {Schmalzl}}, \bibinfo {author} {\bibfnamefont {S.}~\bibnamefont {Raymond}}, \bibinfo {author} {\bibfnamefont {M.}~\bibnamefont {dos Santos~Dias}}, \bibinfo {author} {\bibfnamefont {J.}~\bibnamefont {Persson}}, \bibinfo {author} {\bibfnamefont {N.}~\bibnamefont {Marzari}}, \bibinfo {author} {\bibfnamefont {S.}~\bibnamefont {Bl\"ugel}}, \bibinfo {author} {\bibfnamefont {S.}~\bibnamefont {Lounis}},\ and\ \bibinfo {author} {\bibfnamefont {T.}~\bibnamefont {Br\"uckel}},\ }\bibfield  {title} {\bibinfo {title} {Complex magnetic structure and spin waves of the noncollinear antiferromagnet {${\mathrm{Mn}}_{5}{\mathrm{Si}}_{3}$}},\ }\href {https://doi.org/10.1103/PhysRevB.105.104404} {\bibfield  {journal} {\bibinfo  {journal} {Phys. Rev. B}\ }\textbf {\bibinfo {volume} {105}},\ \bibinfo {pages} {104404} (\bibinfo
  {year} {2022})}\BibitemShut {NoStop}%
\bibitem [{\citenamefont {Harker}(1981)}]{Harker1981}%
  \BibitemOpen
  \bibfield  {author} {\bibinfo {author} {\bibfnamefont {D.}~\bibnamefont {Harker}},\ }\bibfield  {title} {\bibinfo {title} {{The three-colored three-dimensional space groups}},\ }\href {https://doi.org/10.1107/S0567739481000697} {\bibfield  {journal} {\bibinfo  {journal} {Acta Crystallogr. A}\ }\textbf {\bibinfo {volume} {37}},\ \bibinfo {pages} {286} (\bibinfo {year} {1981})}\BibitemShut {NoStop}%
\bibitem [{\citenamefont {Lifshitz}(1997)}]{Lifshitz-RMP}%
  \BibitemOpen
  \bibfield  {author} {\bibinfo {author} {\bibfnamefont {R.}~\bibnamefont {Lifshitz}},\ }\bibfield  {title} {\bibinfo {title} {Theory of color symmetry for periodic and quasiperiodic crystals},\ }\href {https://doi.org/10.1103/RevModPhys.69.1181} {\bibfield  {journal} {\bibinfo  {journal} {Rev. Mod. Phys.}\ }\textbf {\bibinfo {volume} {69}},\ \bibinfo {pages} {1181} (\bibinfo {year} {1997})}\BibitemShut {NoStop}%
\bibitem [{\citenamefont {Jin}\ and\ \citenamefont {Zhou}(2025)}]{Jin2025}%
  \BibitemOpen
  \bibfield  {author} {\bibinfo {author} {\bibfnamefont {J.-T.}\ \bibnamefont {Jin}}\ and\ \bibinfo {author} {\bibfnamefont {Y.}~\bibnamefont {Zhou}},\ }\bibfield  {title} {\bibinfo {title} {Phenomenological {Ginzburg-Landau} theory for {triple-Q} magnetic orders on a hexagonal lattice},\ }\href {https://doi.org/10.1103/nvnm-rtwd} {\bibfield  {journal} {\bibinfo  {journal} {Phys. Rev. B}\ }\textbf {\bibinfo {volume} {112}},\ \bibinfo {pages} {224434} (\bibinfo {year} {2025})}\BibitemShut {NoStop}%
\bibitem [{\citenamefont {Mendoza-Estrada}\ \emph {et~al.}(2025)\citenamefont {Mendoza-Estrada}, \citenamefont {Gonz\'alez-Hern\'andez}, \citenamefont {Uribe},\ and\ \citenamefont {\ifmmode~\check{S}\else \v{S}\fi{}mejkal}}]{MendozaEstrada2025}%
  \BibitemOpen
  \bibfield  {author} {\bibinfo {author} {\bibfnamefont {V.}~\bibnamefont {Mendoza-Estrada}}, \bibinfo {author} {\bibfnamefont {R.}~\bibnamefont {Gonz\'alez-Hern\'andez}}, \bibinfo {author} {\bibfnamefont {B.}~\bibnamefont {Uribe}},\ and\ \bibinfo {author} {\bibfnamefont {L.}~\bibnamefont {\ifmmode~\check{S}\else \v{S}\fi{}mejkal}},\ }\bibfield  {title} {\bibinfo {title} {Eightfold degenerate {Dirac} nodal line in the collinear antiferromagnet {${\mathrm{Mn}}_{5}{\mathrm{Si}}_{3}$}},\ }\href {https://doi.org/10.1103/PhysRevB.111.085147} {\bibfield  {journal} {\bibinfo  {journal} {Phys. Rev. B}\ }\textbf {\bibinfo {volume} {111}},\ \bibinfo {pages} {085147} (\bibinfo {year} {2025})}\BibitemShut {NoStop}%
\bibitem [{\citenamefont {Biniskos}\ \emph {et~al.}(2018)\citenamefont {Biniskos}, \citenamefont {Schmalzl}, \citenamefont {Raymond}, \citenamefont {Petit}, \citenamefont {Steffens}, \citenamefont {Persson},\ and\ \citenamefont {Br\"uckel}}]{Biniskos2018}%
  \BibitemOpen
  \bibfield  {author} {\bibinfo {author} {\bibfnamefont {N.}~\bibnamefont {Biniskos}}, \bibinfo {author} {\bibfnamefont {K.}~\bibnamefont {Schmalzl}}, \bibinfo {author} {\bibfnamefont {S.}~\bibnamefont {Raymond}}, \bibinfo {author} {\bibfnamefont {S.}~\bibnamefont {Petit}}, \bibinfo {author} {\bibfnamefont {P.}~\bibnamefont {Steffens}}, \bibinfo {author} {\bibfnamefont {J.}~\bibnamefont {Persson}},\ and\ \bibinfo {author} {\bibfnamefont {T.}~\bibnamefont {Br\"uckel}},\ }\bibfield  {title} {\bibinfo {title} {Spin fluctuations drive the inverse magnetocaloric effect in {${\mathrm{Mn}}_{5}{\mathrm{Si}}_{3}$}},\ }\href {https://doi.org/10.1103/PhysRevLett.120.257205} {\bibfield  {journal} {\bibinfo  {journal} {Phys. Rev. Lett.}\ }\textbf {\bibinfo {volume} {120}},\ \bibinfo {pages} {257205} (\bibinfo {year} {2018})}\BibitemShut {NoStop}%
\bibitem [{\citenamefont {Biniskos}\ \emph {et~al.}(2023)\citenamefont {Biniskos}, \citenamefont {dos Santos}, \citenamefont {dos Santos~Dias}, \citenamefont {Raymond}, \citenamefont {Schmalzl}, \citenamefont {Steffens}, \citenamefont {Persson}, \citenamefont {Marzari}, \citenamefont {Blügel}, \citenamefont {Lounis},\ and\ \citenamefont {Brückel}}]{Biniskos2023}%
  \BibitemOpen
  \bibfield  {author} {\bibinfo {author} {\bibfnamefont {N.}~\bibnamefont {Biniskos}}, \bibinfo {author} {\bibfnamefont {F.~J.}\ \bibnamefont {dos Santos}}, \bibinfo {author} {\bibfnamefont {M.}~\bibnamefont {dos Santos~Dias}}, \bibinfo {author} {\bibfnamefont {S.}~\bibnamefont {Raymond}}, \bibinfo {author} {\bibfnamefont {K.}~\bibnamefont {Schmalzl}}, \bibinfo {author} {\bibfnamefont {P.}~\bibnamefont {Steffens}}, \bibinfo {author} {\bibfnamefont {J.}~\bibnamefont {Persson}}, \bibinfo {author} {\bibfnamefont {N.}~\bibnamefont {Marzari}}, \bibinfo {author} {\bibfnamefont {S.}~\bibnamefont {Blügel}}, \bibinfo {author} {\bibfnamefont {S.}~\bibnamefont {Lounis}},\ and\ \bibinfo {author} {\bibfnamefont {T.}~\bibnamefont {Brückel}},\ }\bibfield  {title} {\bibinfo {title} {An overview of the spin dynamics of antiferromagnetic {Mn$_5$Si$_3$}},\ }\href {https://doi.org/10.1063/5.0156028} {\bibfield  {journal} {\bibinfo  {journal} {APL Mater.}\ }\textbf {\bibinfo {volume} {11}},\ \bibinfo {pages} {081103} (\bibinfo
  {year} {2023})}\BibitemShut {NoStop}%
\bibitem [{\citenamefont {Todoroki}\ and\ \citenamefont {Miyashita}(2004)}]{TodorokiMiyashita2004}%
  \BibitemOpen
  \bibfield  {author} {\bibinfo {author} {\bibfnamefont {N.}~\bibnamefont {Todoroki}}\ and\ \bibinfo {author} {\bibfnamefont {S.}~\bibnamefont {Miyashita}},\ }\bibfield  {title} {\bibinfo {title} {Ordered phases and phase transitions in the stacked triangular antiferromagnet {CsCoCl$_3$} and {CsCoBr$_3$}},\ }\href {https://doi.org/10.1143/JPSJ.73.412} {\bibfield  {journal} {\bibinfo  {journal} {J. Phys. Soc. Japan}\ }\textbf {\bibinfo {volume} {73}},\ \bibinfo {pages} {412} (\bibinfo {year} {2004})}\BibitemShut {NoStop}%
\bibitem [{\citenamefont {McClarty}\ and\ \citenamefont {Rau}(2024)}]{McClarty2024}%
  \BibitemOpen
  \bibfield  {author} {\bibinfo {author} {\bibfnamefont {P.~A.}\ \bibnamefont {McClarty}}\ and\ \bibinfo {author} {\bibfnamefont {J.~G.}\ \bibnamefont {Rau}},\ }\bibfield  {title} {\bibinfo {title} {Landau theory of altermagnetism},\ }\href {https://doi.org/10.1103/PhysRevLett.132.176702} {\bibfield  {journal} {\bibinfo  {journal} {Phys. Rev. Lett.}\ }\textbf {\bibinfo {volume} {132}},\ \bibinfo {pages} {176702} (\bibinfo {year} {2024})}\BibitemShut {NoStop}%
\bibitem [{\citenamefont {Schiff}\ \emph {et~al.}(2025)\citenamefont {Schiff}, \citenamefont {McClarty}, \citenamefont {Rau},\ and\ \citenamefont {Romh\'anyi}}]{Schiff2025}%
  \BibitemOpen
  \bibfield  {author} {\bibinfo {author} {\bibfnamefont {H.}~\bibnamefont {Schiff}}, \bibinfo {author} {\bibfnamefont {P.}~\bibnamefont {McClarty}}, \bibinfo {author} {\bibfnamefont {J.~G.}\ \bibnamefont {Rau}},\ and\ \bibinfo {author} {\bibfnamefont {J.}~\bibnamefont {Romh\'anyi}},\ }\bibfield  {title} {\bibinfo {title} {Collinear altermagnets and their {Landau} theories},\ }\href {https://doi.org/10.1103/q44z-ynbr} {\bibfield  {journal} {\bibinfo  {journal} {Phys. Rev. Res.}\ }\textbf {\bibinfo {volume} {7}},\ \bibinfo {pages} {033301} (\bibinfo {year} {2025})}\BibitemShut {NoStop}%
\bibitem [{\citenamefont {Staunton}\ \emph {et~al.}(1986)\citenamefont {Staunton}, \citenamefont {Gyorffy}, \citenamefont {Stocks},\ and\ \citenamefont {Wadsworth}}]{Staunton1986}%
  \BibitemOpen
  \bibfield  {author} {\bibinfo {author} {\bibfnamefont {J.}~\bibnamefont {Staunton}}, \bibinfo {author} {\bibfnamefont {B.~L.}\ \bibnamefont {Gyorffy}}, \bibinfo {author} {\bibfnamefont {G.~M.}\ \bibnamefont {Stocks}},\ and\ \bibinfo {author} {\bibfnamefont {J.}~\bibnamefont {Wadsworth}},\ }\bibfield  {title} {\bibinfo {title} {The static, paramagnetic, spin susceptibility of metals at finite temperatures},\ }\href {https://doi.org/10.1088/0305-4608/16/11/016} {\bibfield  {journal} {\bibinfo  {journal} {J. Phys. F: Met. Phys.}\ }\textbf {\bibinfo {volume} {16}},\ \bibinfo {pages} {1761} (\bibinfo {year} {1986})}\BibitemShut {NoStop}%
\bibitem [{\citenamefont {Liechtenstein}\ \emph {et~al.}(1987)\citenamefont {Liechtenstein}, \citenamefont {Katsnelson}, \citenamefont {Antropov},\ and\ \citenamefont {Gubanov}}]{LKAG}%
  \BibitemOpen
  \bibfield  {author} {\bibinfo {author} {\bibfnamefont {A.}~\bibnamefont {Liechtenstein}}, \bibinfo {author} {\bibfnamefont {M.}~\bibnamefont {Katsnelson}}, \bibinfo {author} {\bibfnamefont {V.}~\bibnamefont {Antropov}},\ and\ \bibinfo {author} {\bibfnamefont {V.}~\bibnamefont {Gubanov}},\ }\bibfield  {title} {\bibinfo {title} {Local spin density functional approach to the theory of exchange interactions in ferromagnetic metals and alloys},\ }\href {https://doi.org/https://doi.org/10.1016/0304-8853(87)90721-9} {\bibfield  {journal} {\bibinfo  {journal} {J. Magn. Magn. Mater.}\ }\textbf {\bibinfo {volume} {67}},\ \bibinfo {pages} {65} (\bibinfo {year} {1987})}\BibitemShut {NoStop}%
\bibitem [{\citenamefont {Ondr\'a\ifmmode~\check{c}\else \v{c}\fi{}ek}\ \emph {et~al.}(2010)\citenamefont {Ondr\'a\ifmmode~\check{c}\else \v{c}\fi{}ek}, \citenamefont {Bengone}, \citenamefont {Kudrnovsk\'y}, \citenamefont {Drchal}, \citenamefont {M\'aca},\ and\ \citenamefont {Turek}}]{Ondracek2010}%
  \BibitemOpen
  \bibfield  {author} {\bibinfo {author} {\bibfnamefont {M.}~\bibnamefont {Ondr\'a\ifmmode~\check{c}\else \v{c}\fi{}ek}}, \bibinfo {author} {\bibfnamefont {O.}~\bibnamefont {Bengone}}, \bibinfo {author} {\bibfnamefont {J.}~\bibnamefont {Kudrnovsk\'y}}, \bibinfo {author} {\bibfnamefont {V.}~\bibnamefont {Drchal}}, \bibinfo {author} {\bibfnamefont {F.}~\bibnamefont {M\'aca}},\ and\ \bibinfo {author} {\bibfnamefont {I.}~\bibnamefont {Turek}},\ }\bibfield  {title} {\bibinfo {title} {Magnetic phase stability of monolayers: {Fe} on a {${\text{Ta}}_{x}{\text{W}}_{1\ensuremath{-}x}(001)$} random alloy as a case study},\ }\href {https://doi.org/10.1103/PhysRevB.81.064410} {\bibfield  {journal} {\bibinfo  {journal} {Phys. Rev. B}\ }\textbf {\bibinfo {volume} {81}},\ \bibinfo {pages} {064410} (\bibinfo {year} {2010})}\BibitemShut {NoStop}%
\bibitem [{\citenamefont {Oguchi}\ \emph {et~al.}(1983)\citenamefont {Oguchi}, \citenamefont {Terakura},\ and\ \citenamefont {Hamada}}]{OguchiTerakuraHamada1983}%
  \BibitemOpen
  \bibfield  {author} {\bibinfo {author} {\bibfnamefont {T.}~\bibnamefont {Oguchi}}, \bibinfo {author} {\bibfnamefont {K.}~\bibnamefont {Terakura}},\ and\ \bibinfo {author} {\bibfnamefont {N.}~\bibnamefont {Hamada}},\ }\bibfield  {title} {\bibinfo {title} {Magnetism of iron above the {Curie} temperature},\ }\href {https://doi.org/10.1088/0305-4608/13/1/018} {\bibfield  {journal} {\bibinfo  {journal} {J. Phys. F: Metal Phys.}\ }\textbf {\bibinfo {volume} {13}},\ \bibinfo {pages} {145} (\bibinfo {year} {1983})}\BibitemShut {NoStop}%
\bibitem [{\citenamefont {Gy{\"o}rffy}\ \emph {et~al.}(1985)\citenamefont {Gy{\"o}rffy}, \citenamefont {Pindor}, \citenamefont {Staunton}, \citenamefont {Stocks},\ and\ \citenamefont {Winter}}]{GyorffyPindorStauntonStocksWinter1985}%
  \BibitemOpen
  \bibfield  {author} {\bibinfo {author} {\bibfnamefont {B.~L.}\ \bibnamefont {Gy{\"o}rffy}}, \bibinfo {author} {\bibfnamefont {A.~J.}\ \bibnamefont {Pindor}}, \bibinfo {author} {\bibfnamefont {J.}~\bibnamefont {Staunton}}, \bibinfo {author} {\bibfnamefont {G.~M.}\ \bibnamefont {Stocks}},\ and\ \bibinfo {author} {\bibfnamefont {H.}~\bibnamefont {Winter}},\ }\bibfield  {title} {\bibinfo {title} {A first-principles theory of ferromagnetic phase transitions in metals},\ }\href {https://doi.org/10.1088/0305-4608/15/6/018} {\bibfield  {journal} {\bibinfo  {journal} {J. Phys. F: Metal Phys.}\ }\textbf {\bibinfo {volume} {15}},\ \bibinfo {pages} {1337} (\bibinfo {year} {1985})}\BibitemShut {NoStop}%
\bibitem [{\citenamefont {Turek}\ \emph {et~al.}(1997)\citenamefont {Turek}, \citenamefont {Drchal}, \citenamefont {Kudrnovsk{\'y}}, \citenamefont {{\v S}ob},\ and\ \citenamefont {Weinberger}}]{Turek1997}%
  \BibitemOpen
  \bibfield  {author} {\bibinfo {author} {\bibfnamefont {I.}~\bibnamefont {Turek}}, \bibinfo {author} {\bibfnamefont {V.}~\bibnamefont {Drchal}}, \bibinfo {author} {\bibfnamefont {J.}~\bibnamefont {Kudrnovsk{\'y}}}, \bibinfo {author} {\bibfnamefont {M.}~\bibnamefont {{\v S}ob}},\ and\ \bibinfo {author} {\bibfnamefont {P.}~\bibnamefont {Weinberger}},\ }\href {https://doi.org/10.1007/978-1-4615-6255-9} {\emph {\bibinfo {title} {Electronic Structure of Disordered Alloys, Surfaces and Interfaces}}}\ (\bibinfo  {publisher} {Springer},\ \bibinfo {address} {New York},\ \bibinfo {year} {1997})\BibitemShut {NoStop}%
\bibitem [{\citenamefont {Andersen}(1975)}]{Andersen1975}%
  \BibitemOpen
  \bibfield  {author} {\bibinfo {author} {\bibfnamefont {O.~K.}\ \bibnamefont {Andersen}},\ }\bibfield  {title} {\bibinfo {title} {Linear methods in band theory},\ }\href {https://doi.org/10.1103/PhysRevB.12.3060} {\bibfield  {journal} {\bibinfo  {journal} {Phys. Rev. B}\ }\textbf {\bibinfo {volume} {12}},\ \bibinfo {pages} {3060} (\bibinfo {year} {1975})}\BibitemShut {NoStop}%
\bibitem [{\citenamefont {Ke}\ \emph {et~al.}(2013)\citenamefont {Ke}, \citenamefont {Belashchenko}, \citenamefont {van Schilfgaarde}, \citenamefont {Kotani},\ and\ \citenamefont {Antropov}}]{Ke2013}%
  \BibitemOpen
  \bibfield  {author} {\bibinfo {author} {\bibfnamefont {L.}~\bibnamefont {Ke}}, \bibinfo {author} {\bibfnamefont {K.~D.}\ \bibnamefont {Belashchenko}}, \bibinfo {author} {\bibfnamefont {M.}~\bibnamefont {van Schilfgaarde}}, \bibinfo {author} {\bibfnamefont {T.}~\bibnamefont {Kotani}},\ and\ \bibinfo {author} {\bibfnamefont {V.~P.}\ \bibnamefont {Antropov}},\ }\bibfield  {title} {\bibinfo {title} {Effects of alloying and strain on the magnetic properties of {Fe${}_{16}$N${}_{2}$}},\ }\href {https://doi.org/10.1103/PhysRevB.88.024404} {\bibfield  {journal} {\bibinfo  {journal} {Phys. Rev. B}\ }\textbf {\bibinfo {volume} {88}},\ \bibinfo {pages} {024404} (\bibinfo {year} {2013})}\BibitemShut {NoStop}%
\bibitem [{\citenamefont {Pashov}\ \emph {et~al.}(2020)\citenamefont {Pashov}, \citenamefont {Acharya}, \citenamefont {Lambrecht}, \citenamefont {Jackson}, \citenamefont {Belashchenko}, \citenamefont {Chantis}, \citenamefont {Jamet},\ and\ \citenamefont {{van Schilfgaarde}}}]{Pashov2020:CPC}%
  \BibitemOpen
  \bibfield  {author} {\bibinfo {author} {\bibfnamefont {D.}~\bibnamefont {Pashov}}, \bibinfo {author} {\bibfnamefont {S.}~\bibnamefont {Acharya}}, \bibinfo {author} {\bibfnamefont {W.~R.}\ \bibnamefont {Lambrecht}}, \bibinfo {author} {\bibfnamefont {J.}~\bibnamefont {Jackson}}, \bibinfo {author} {\bibfnamefont {K.~D.}\ \bibnamefont {Belashchenko}}, \bibinfo {author} {\bibfnamefont {A.}~\bibnamefont {Chantis}}, \bibinfo {author} {\bibfnamefont {F.}~\bibnamefont {Jamet}},\ and\ \bibinfo {author} {\bibfnamefont {M.}~\bibnamefont {{van Schilfgaarde}}},\ }\bibfield  {title} {\bibinfo {title} {Questaal: A package of electronic structure methods based on the linear muffin-tin orbital technique},\ }\href {https://doi.org/https://doi.org/10.1016/j.cpc.2019.107065} {\bibfield  {journal} {\bibinfo  {journal} {Comput. Phys. Commun.}\ }\textbf {\bibinfo {volume} {249}},\ \bibinfo {pages} {107065} (\bibinfo {year} {2020})}\BibitemShut {NoStop}%
\bibitem [{\citenamefont {Lander}\ and\ \citenamefont {Brown}(1967)}]{Lander01091967}%
  \BibitemOpen
  \bibfield  {author} {\bibinfo {author} {\bibfnamefont {G.~H.}\ \bibnamefont {Lander}}\ and\ \bibinfo {author} {\bibfnamefont {P.~J.}\ \bibnamefont {Brown}},\ }\bibfield  {title} {\bibinfo {title} {Electron density distribution in the alloy {Mn$_5$Si$_3$}},\ }\href {https://doi.org/10.1080/14786436708220862} {\bibfield  {journal} {\bibinfo  {journal} {Philos. Mag.}\ }\textbf {\bibinfo {volume} {16}},\ \bibinfo {pages} {521} (\bibinfo {year} {1967})}\BibitemShut {NoStop}%
\bibitem [{\citenamefont {von Barth}\ and\ \citenamefont {Hedin}(1972)}]{vonBarthHedin1972}%
  \BibitemOpen
  \bibfield  {author} {\bibinfo {author} {\bibfnamefont {U.}~\bibnamefont {von Barth}}\ and\ \bibinfo {author} {\bibfnamefont {L.}~\bibnamefont {Hedin}},\ }\bibfield  {title} {\bibinfo {title} {A local exchange-correlation potential for the spin polarized case. {I}},\ }\href {https://doi.org/10.1088/0022-3719/5/13/012} {\bibfield  {journal} {\bibinfo  {journal} {J. Phys. C: Solid State Phys.}\ }\textbf {\bibinfo {volume} {5}},\ \bibinfo {pages} {1629} (\bibinfo {year} {1972})}\BibitemShut {NoStop}%
\bibitem [{\citenamefont {Perdew}\ \emph {et~al.}(1996)\citenamefont {Perdew}, \citenamefont {Burke},\ and\ \citenamefont {Ernzerhof}}]{PerdewBurkeErnzerhof1996}%
  \BibitemOpen
  \bibfield  {author} {\bibinfo {author} {\bibfnamefont {J.~P.}\ \bibnamefont {Perdew}}, \bibinfo {author} {\bibfnamefont {K.}~\bibnamefont {Burke}},\ and\ \bibinfo {author} {\bibfnamefont {M.}~\bibnamefont {Ernzerhof}},\ }\bibfield  {title} {\bibinfo {title} {Generalized gradient approximation made simple},\ }\href {https://doi.org/10.1103/PhysRevLett.77.3865} {\bibfield  {journal} {\bibinfo  {journal} {Phys. Rev. Lett.}\ }\textbf {\bibinfo {volume} {77}},\ \bibinfo {pages} {3865} (\bibinfo {year} {1996})}\BibitemShut {NoStop}%
\bibitem [{\citenamefont {Ortenzi}\ \emph {et~al.}(2012)\citenamefont {Ortenzi}, \citenamefont {Mazin}, \citenamefont {Blaha},\ and\ \citenamefont {Boeri}}]{Ortenzi2012}%
  \BibitemOpen
  \bibfield  {author} {\bibinfo {author} {\bibfnamefont {L.}~\bibnamefont {Ortenzi}}, \bibinfo {author} {\bibfnamefont {I.~I.}\ \bibnamefont {Mazin}}, \bibinfo {author} {\bibfnamefont {P.}~\bibnamefont {Blaha}},\ and\ \bibinfo {author} {\bibfnamefont {L.}~\bibnamefont {Boeri}},\ }\bibfield  {title} {\bibinfo {title} {Accounting for spin fluctuations beyond local spin density approximation in the density functional theory},\ }\href {https://doi.org/10.1103/PhysRevB.86.064437} {\bibfield  {journal} {\bibinfo  {journal} {Phys. Rev. B}\ }\textbf {\bibinfo {volume} {86}},\ \bibinfo {pages} {064437} (\bibinfo {year} {2012})}\BibitemShut {NoStop}%
\bibitem [{\citenamefont {de~Jong}\ \emph {et~al.}(2015)\citenamefont {de~Jong}, \citenamefont {Chen}, \citenamefont {Angsten}, \citenamefont {Jain}, \citenamefont {Notestine}, \citenamefont {Gamst}, \citenamefont {Sluiter}, \citenamefont {Ande}, \citenamefont {van~der Zwaag}, \citenamefont {Plata}, \citenamefont {Toher}, \citenamefont {Curtarolo}, \citenamefont {Ceder}, \citenamefont {Persson},\ and\ \citenamefont {Asta}}]{deJong2015Elastic}%
  \BibitemOpen
  \bibfield  {author} {\bibinfo {author} {\bibfnamefont {M.}~\bibnamefont {de~Jong}}, \bibinfo {author} {\bibfnamefont {W.}~\bibnamefont {Chen}}, \bibinfo {author} {\bibfnamefont {T.}~\bibnamefont {Angsten}}, \bibinfo {author} {\bibfnamefont {A.}~\bibnamefont {Jain}}, \bibinfo {author} {\bibfnamefont {R.}~\bibnamefont {Notestine}}, \bibinfo {author} {\bibfnamefont {A.}~\bibnamefont {Gamst}}, \bibinfo {author} {\bibfnamefont {M.}~\bibnamefont {Sluiter}}, \bibinfo {author} {\bibfnamefont {C.~K.}\ \bibnamefont {Ande}}, \bibinfo {author} {\bibfnamefont {S.}~\bibnamefont {van~der Zwaag}}, \bibinfo {author} {\bibfnamefont {J.~J.}\ \bibnamefont {Plata}}, \bibinfo {author} {\bibfnamefont {C.}~\bibnamefont {Toher}}, \bibinfo {author} {\bibfnamefont {S.}~\bibnamefont {Curtarolo}}, \bibinfo {author} {\bibfnamefont {G.}~\bibnamefont {Ceder}}, \bibinfo {author} {\bibfnamefont {K.~A.}\ \bibnamefont {Persson}},\ and\ \bibinfo {author} {\bibfnamefont {M.}~\bibnamefont {Asta}},\ }\bibfield  {title} {\bibinfo {title} {Charting
  the complete elastic properties of inorganic crystalline compounds},\ }\href {https://doi.org/10.1038/sdata.2015.9} {\bibfield  {journal} {\bibinfo  {journal} {Sci. Data}\ }\textbf {\bibinfo {volume} {2}},\ \bibinfo {pages} {150009} (\bibinfo {year} {2015})}\BibitemShut {NoStop}%
\bibitem [{\citenamefont {de~Jong}\ \emph {et~al.}(2016)\citenamefont {de~Jong}, \citenamefont {Chen}, \citenamefont {Angsten} \emph {et~al.}}]{deJong2016ElasticData}%
  \BibitemOpen
  \bibfield  {author} {\bibinfo {author} {\bibfnamefont {M.}~\bibnamefont {de~Jong}}, \bibinfo {author} {\bibfnamefont {W.}~\bibnamefont {Chen}}, \bibinfo {author} {\bibfnamefont {T.}~\bibnamefont {Angsten}}, \emph {et~al.},\ }\bibfield  {title} {\bibinfo {title} {Data from: Charting the complete elastic properties of inorganic crystalline compounds},\ }\href {https://doi.org/10.5061/dryad.h505v} {10.5061/dryad.h505v} (\bibinfo {year} {2016})\BibitemShut {NoStop}%
\bibitem [{\citenamefont {Skubic}\ \emph {et~al.}(2008)\citenamefont {Skubic}, \citenamefont {Hellsvik}, \citenamefont {Nordstr{\"o}m},\ and\ \citenamefont {Eriksson}}]{UppASD}%
  \BibitemOpen
  \bibfield  {author} {\bibinfo {author} {\bibfnamefont {B.}~\bibnamefont {Skubic}}, \bibinfo {author} {\bibfnamefont {J.}~\bibnamefont {Hellsvik}}, \bibinfo {author} {\bibfnamefont {L.}~\bibnamefont {Nordstr{\"o}m}},\ and\ \bibinfo {author} {\bibfnamefont {O.}~\bibnamefont {Eriksson}},\ }\bibfield  {title} {\bibinfo {title} {A method for atomistic spin dynamics simulations: implementation and examples},\ }\href {https://doi.org/10.1088/0953-8984/20/31/315203} {\bibfield  {journal} {\bibinfo  {journal} {J. Phys.: Condens. Matter}\ }\textbf {\bibinfo {volume} {20}},\ \bibinfo {pages} {315203} (\bibinfo {year} {2008})}\BibitemShut {NoStop}%
\end{thebibliography}

%

\end{document}